\documentclass[a4paper,11pt]{article}
\usepackage{jcappub} 
\usepackage{lineno}
\usepackage{comment}
\usepackage{graphicx}
\usepackage{setspace,caption}
\usepackage{array,amsmath,slashed,amsthm,amsfonts,amssymb,xcolor,tikz,float}
\usepackage{booktabs,multirow}
\newcolumntype{C}[1]{>{\centering\arraybackslash}m{#1}}
    
\newcommand{\AddrIACS}{%
	School of Physical Sciences, Indian Association for the	Cultivation of Science, Jadavpur, Kolkata 700032, India}

\usepackage[normalem]{ulem}

\title{Diffuse Supernova Neutrinos with Secret Neutrino Interactions}

\author[1]{Praveen Bharadwaj,}\note{https://orcid.org/0000-0001-7309-101X}
\author[2]{Utpal Chattopadhyay,}\note{https://orcid.org/0000-0002-6824-9465}
\author[3]{Dilip Kumar Ghosh,}\note{https://orcid.org/0000-0003-1764-1632}
\author[4]{Arnab Sarker}\note{https://orcid.org/0000-0002-5211-0338}
\affiliation{\AddrIACS}
\emailAdd{spspb3454@iacs.res.in}
\emailAdd{tpuc@iacs.res.in}
\emailAdd{tpdkg@iacs.res.in}
\emailAdd{spsas3453@iacs.res.in}

\abstract{The Diffuse Supernova Neutrino Background (DSNB), an isotropic flux arising from the cumulative neutrino emission of all stellar core-collapse events throughout cosmic history, is expected to be detected by next-generation neutrino observatories. As DSNB neutrinos propagate over cosmological distances through the cosmic neutrino background (C$\nu$B), they may undergo non-standard neutrino self-interactions ($\nu$SI), leaving distinct spectral imprints on the observed flux. In this work, we investigate the impact of scalar ($\phi$)-mediated $\nu$SI on the DSNB within a full three-flavor framework that retains the complete PMNS structure. We consider four representative flavor-diagonal coupling structures---universal, $e$-, $\mu$-, and $\tau$-specific. The resonant scattering $\nu_i\nu_k\to\phi\to\nu_j\nu_l$ off the lightest, relativistic C$\nu$B state produces broad spectral depletion whose pattern depends on the coupling structure and the neutrino mass ordering, generating distinctive signatures across the six flavor fluxes. We compute the resulting event spectra at JUNO, Hyper-Kamiokande with gadolinium loading, and DUNE, and derive projected $3\sigma$ sensitivities in the $(m_{\phi},~g)$ parameter plane. We find that these experiments can probe couplings as low as $g\sim10^{-8}$ for $m_\phi\sim100$--$300$~eV, surpassing existing bounds by up to a few orders of magnitude in the sub-100~eV mass range. Moreover, unlike the flavor-blind cosmological and supernova bounds, the DSNB sensitivity is flavor-discriminating, offering a unique opportunity to identify the underlying flavor structure of $\nu$SI in the event of a detection.}

\begin{document}
\maketitle
\flushbottom

\section{\label{sec:intro}Introduction}
The Diffuse Supernova Neutrino Background (DSNB) is an isotropic and time-independent flux of neutrinos and antineutrinos emitted by the cumulative population of stellar core collapses throughout cosmic history~\cite{Lunardini:2010ab,Beacom:2010kk,Ando:2023fcc}. First proposed in the 1980s~\cite{Bisnovatyi-Kogan:1982oyy,Krauss:1983zn}, the DSNB is yet to be experimentally confirmed. However, the limits from Super-Kamiokande~\cite{Super-Kamiokande:2021jaq,Super-Kamiokande:2011lwo}, particularly those from the latest gadolinium-doped phase~\cite{Super-Kamiokande:2025sxh}, are only a few factors above the theoretical predictions, indicating that the sensitivity required for discovery is within reach of present and upcoming neutrino detectors. 
The successful detection of the DSNB would represent a landmark achievement in neutrino astronomy, providing insight into poorly understood aspects of core-collapse physics, including the explosion mechanism, the cosmic core-collapse rate, the nuclear equation of state, and the fraction of collapses that form black holes~\cite{Nakazato:2013maa,Mirizzi:2015eza,Horiuchi:2018ofe,Moller:2018kpn,Warren:2019lgb,Burrows:2020qrp,Suliga:2021hek}, as well as information on neutrino properties integrated over cosmic history~\cite{Chakraborty:2008zp,Lunardini:2012ne,DeGouvea:2020ang,deGouvea:2022dtw,Martinez-Mirave:2024hfd}. On the other hand, a null result or a measured spectrum that deviates from theoretical predictions could signal the presence of new physics, such as sterile neutrinos~\cite{Jeong:2018yts,Tang:2020pkp}, non-radiative neutrino decays~\cite{Tabrizi:2020vmo,Ivanez-Ballesteros:2022szu,MacDonald:2024vtw,Ivanez-Ballesteros:2025ojj}, or secret interactions among neutrinos~\cite{Goldberg:2005yw,Baker:2006gm,Creque-Sarbinowski:2020qhz,Das:2022xsz,Akita:2022etk,Wang:2025qap}. 

Secret neutrino interactions, also known as non-standard neutrino self-interactions ($\nu$SI), are a well-motivated extension of the Standard Model (SM) with implications in astrophysics, cosmology, and particle physics (see ref.~\cite{Berryman:2022hds} for a review). Among these motivations is their possible connection to dark matter~\cite{DeGouvea:2019wpf,Lattanzi:2007ux,Frigerio:2011in,Garcia-Cely:2017oco,Cline:2026tkp}, as well as their potential role in addressing the Hubble and $S_8$ tensions~\cite{Kreisch:2019yzn,Blinov:2019gcj,Escudero:2019gvw}.
Perhaps the most compelling motivation is their close relation to neutrino mass generation~\cite{Chikashige:1980ui,Gelmini:1980re,GEORGI1981297}, since non-zero neutrino masses remain the only direct evidence of beyond the Standard Model (BSM) physics~\cite{deGouvea:2016qpx,ParticleDataGroup:2024cfk}. 
In many such models, the same new dynamics that generate neutrino masses also give rise to light scalar or vector mediators, leading to additional neutrino-neutrino interactions with strengths that can exceed the SM weak interaction by several orders of magnitude. Such interactions can affect cosmological perturbations, the propagation of astrophysical neutrinos, and their flavor and transport properties in dense environments, making $\nu$SI a highly compelling avenue for both theoretical and experimental studies.

The search for $\nu$SI spans a diverse range of experimental and observational frontiers. In the terrestrial regime, $\nu$SI are tightly constrained by high-precision laboratory measurements, such as meson and tau decays~\cite{Bilenky:1992xn,Lessa:2007up,Pasquini:2015fjv,Lyu:2020lps,PIENU:2021clt,Brdar:2020nbj}, as well as lepton number violating searches in neutrinoless double beta decay ($0\nu\beta\beta$)~\cite{Kharusi:2021jez,KamLAND-Zen:2012uen,Blum:2018ljv,Deppisch:2020sqh}.
However, observing the effects of $\nu$SI is challenging, as it typically requires either intense neutrino fluxes or environments with extremely high neutrino densities. As such conditions are difficult to realize on Earth, studies often rely on cosmological constraints~\cite{Li:2023puz,Huang:2017egl,Sandner:2023ptm} and astrophysical systems such as stellar interiors and core-collapse supernovae~\cite{Manohar:1987ec,Kachelriess:2000qc,Farzan:2002wx,Heurtier:2016otg,Fiorillo:2022cdq,Akita:2022etk,Wu:2023twu,Ivanez-Ballesteros:2024nws} which naturally host these extreme conditions.
Beyond these dense astrophysical sources, high-energy astrophysical neutrinos propagating over cosmic baselines provide another powerful probe. As these neutrinos traverse the Universe, their flux could be attenuated by $\nu$SI-induced scattering off the cosmic neutrino background (C$\nu$B)~\cite{Kolb:1987qy,Baker:2006gm,Ng:2014pca,Ioka:2014kca,Ibe:2014pja,Shalgar:2019rqe,Bustamante:2020mep,Creque-Sarbinowski:2020qhz,Esteban:2021tub,Das:2022xsz,Leal:2025eou}~\footnote{Note that if the underlying $\nu$SI mediator also couples to the dark sector, attenuation could similarly occur via neutrino scattering off dark matter particles within galactic halos~\cite{Farzan:2014gza,Balantekin:2023jlg,Leal:2025eou}.}. 
Furthermore, it was recently noted in~\cite{Doring:2023vmk,Wang:2025qap,Machado:2025ltu} that this attenuation can be significantly amplified through resonant \(s\)-channel scattering ($\nu\nu\to\phi\to\nu\nu$), if the lightest neutrino species in C$\nu$B is still relativistic today, which is consistent with current experimental constraints~\footnote{This has inspired a wide range of `minimal' neutrino mass models where the lightest neutrino remains massless, see for example~\cite{Malinsky:2009df,Gavela:2009cd,Khan:2012zw,Abada:2014vea,Batra:2023ssq}.}. The thermal momentum spread of the relic neutrinos causes the resonance condition to be met over a broad range of incoming neutrino energies, producing a wide absorption feature in the spectrum, in contrast to the sharp spectral dip at $E_{\rm res}\simeq m_\phi^2/2m_\nu$ that characterizes the non-relativistic case.

In this work, we evaluate how $\nu$SI-mediated scattering of DSNB neutrinos off the relic C$\nu$B distorts the DSNB flux spectra. We focus on the relativistic C$\nu$B scenario, where the resulting broad absorption signature significantly enhances the sensitivity to $\nu$SI, allowing us to probe couplings as small as $\sim10^{-8}$ for mediator masses in the eV--keV range~\cite{Wang:2025qap}. In contrast to earlier analyses, we work within a full three-flavor framework, using flavor-dependent emission spectra from core-collapse simulations. We consider both normal and inverted mass orderings, which affect the predicted spectra through the MSW flavor conversion at the source and through the identity of the relativistic C$\nu$B target state. Additionally, we consider four representative flavor structures of the $\nu$SI coupling---universal, $e$-specific, $\mu$-specific, and $\tau$-specific, and show that each induces distinct absorption features in the DSNB spectrum. We present the projected sensitivities at current and next-generation observatories, including the Jiangmen Underground Neutrino Observatory (JUNO)~\cite{JUNO:2015zny,JUNO:2022lpc}, Hyper-Kamiokande with Gadolinium (HK--Gd)~\cite{Hyper-Kamiokande:2018ofw}, and the Deep Underground Neutrino Experiment (DUNE)~\cite{DUNE:2020ypp}. These results provide a complementary probe of the $\nu$SI parameter space, independent of laboratory and cosmological constraints.

The paper is organized as follows. In section~\ref{sec:dsnb}, we discuss the theoretical treatment of the DSNB flux. In section~\ref{sec:nusi}, we provide an overview of neutrino self-interactions and explore how they modify the DSNB spectra. We then discuss the detection prospects of the $\rm\nu SI$-modified DSNB spectra at JUNO, HK, and DUNE in section~\ref{events}. The projected sensitivities are discussed in section~\ref {res}, followed by our conclusions in section~\ref {conc}.

\section{\label{sec:dsnb}The Diffuse Supernova Neutrino Background}
During a typical core-collapse event, an intense burst of neutrinos and antineutrinos is emitted, carrying away nearly all the gravitational binding energy ($E^\text{tot}_\nu\sim3\times10^{53}$ erg) of the newly formed compact remnant---neutron star or black hole~\cite{Mirizzi:2015eza,Burrows:2020qrp}. Although this energy is roughly divided among the six species, their mean energies and luminosities differ, reflecting the flavor-dependent decoupling at the neutrinosphere. 
The energy spectra and relative luminosities of these neutrinos encode information about the core-collapse dynamics and the nature of the resulting compact remnant.
While a galactic core-collapse event would deliver a high-statistics burst, such events are rare ($1.63\pm0.46$ events per century~\cite{Rozwadowska:2020nab}). The DSNB, arising from the cumulative neutrino emission from all past stellar core collapses, thus offers a steady and guaranteed signal within the reach of next-generation neutrino observatories.

In this section, we discuss the theoretical description of the DSNB flux spectrum, which depends primarily on three components: the neutrino emission spectrum per core-collapse event $dN_{\nu_\beta}/dE$, the evolution of the core-collapse rate density $R_\text{CC}(z)$, and the transport of neutrinos across cosmological distances in an expanding Universe. Each component is discussed below.

\subsection{Neutrino Emission Spectra}
The neutrino emission from a core-collapse event depends on the progenitor structure and on the outcome of collapse, namely a successful explosion leaving a neutron star, or a failed explosion resulting in black hole formation. Failed explosions typically feature a longer accretion phase and a harder neutrino spectrum before an abrupt cutoff at black hole formation, whereas successful explosions are dominated by the subsequent proto-neutron star cooling phase, with a softer time-integrated spectrum. Modern multi-progenitor simulations capture this diversity by providing time-dependent luminosities $L_\beta(t)$, mean energies $\langle E_\beta(t)\rangle$, and shape parameters $\alpha_\beta(t)$ for each flavor $\beta=\nu_e,\bar\nu_e,\nu_x$, where $\nu_x$ denotes a representative non-electron flavor or anti-flavor ($\nu_\mu,\bar\nu_\mu,\nu_\tau,\bar\nu_\tau$), so that $L_{\nu_x}$ is the per-species luminosity~\cite{Kresse:2020nto,Horiuchi:2017qja}. The differential number spectrum at a given time is parameterized in the pinched-thermal form as~\footnote{We omit the species label $\beta$, as the same parameterization applies to all flavors.} 
\begin{equation}
\frac{d^2N}{dtdE}(E,t)=\frac{L(t)}{\langle E(t)\rangle}\,\varphi(E,t),
\end{equation}
with the normalized shape function~\cite{Keil:2002in}
\begin{equation}
\varphi(E,t)=\frac{(1+\alpha)^{1+\alpha}}{\Gamma(1+\alpha)}\frac{E^{\alpha}}{\langle E(t)\rangle^{\alpha+1}}\,\exp\left[-(1+\alpha)\frac{E}{\langle E(t)\rangle}\right].
\end{equation}
This form satisfies the following,
\begin{equation}
\int_0^\infty \varphi(E,t)\,dE = 1,
\qquad
\int_0^\infty E\,\varphi(E,t)\,dE = \langle E(t)\rangle.
\end{equation}
The parameter $\alpha$ controls the spectral pinching: larger values correspond to narrower spectra around the mean energy. In practice, $\alpha$ varies with time, progenitor, and flavor, but values in the range $\alpha \sim 2$--$4$ are typical.

\noindent The time-integrated spectrum from a single progenitor is then given by,
\begin{equation}
\left(\frac{dN}{dE}\right)_{\!\rm prog}=\int_0^{t_f} dt\,\frac{d^2N}{dtdE}(E,t).
\end{equation}
For failed explosions, the upper limit $t_f$ is the black hole formation time $t_{\rm BH}$. For successful explosions, $t_f$ is chosen to include the late cooling phase.

The DSNB source spectrum is a population average over all core-collapse progenitors. Since stars are born with different zero-age main sequence (ZAMS) masses according to the initial mass function (IMF) $\psi(M)=dn/dM$, each progenitor model must be weighted by the fraction of stars expected to populate the corresponding mass interval. In practice, the IMF average is approximated by a discrete sum over the progenitor grid:
\begin{align}
    \left\langle\frac{dN}{dE} \right\rangle_{\rm IMF}=\sum_j\frac{\int_{\Delta M_j}\psi(M)dM}{\int_{M_{\rm min}}^{M_{\rm max}}\psi(M)dM}\left(\frac{dN}{dE}\right)_j
\end{align}
where $\Delta M_j$ is the ZAMS-mass interval assigned to the $j$th progenitor model, and $M_{\rm min}=8.7M_\odot$ and $M_{\rm max}=125M_\odot$ are the adopted lower and upper mass limits for core collapse. This weighting accounts for the fact that massive progenitors are intrinsically rarer. For the IMF we adopt the modified Salpeter-A form of ref.~\cite{Baldry:2003xi}, as used in ref.~\cite{Kresse:2020nto}:
\begin{align}\label{eq:imf}
    \psi(M)\propto\begin{cases}
        M^{-1.5}&0.1M_\odot\leq M<0.5M_\odot\\
        M^{-2.35}&M\geq0.5M_\odot\,.
    \end{cases}
\end{align}
\begin{figure}[h]
    \centering    
    \includegraphics[width=\linewidth]{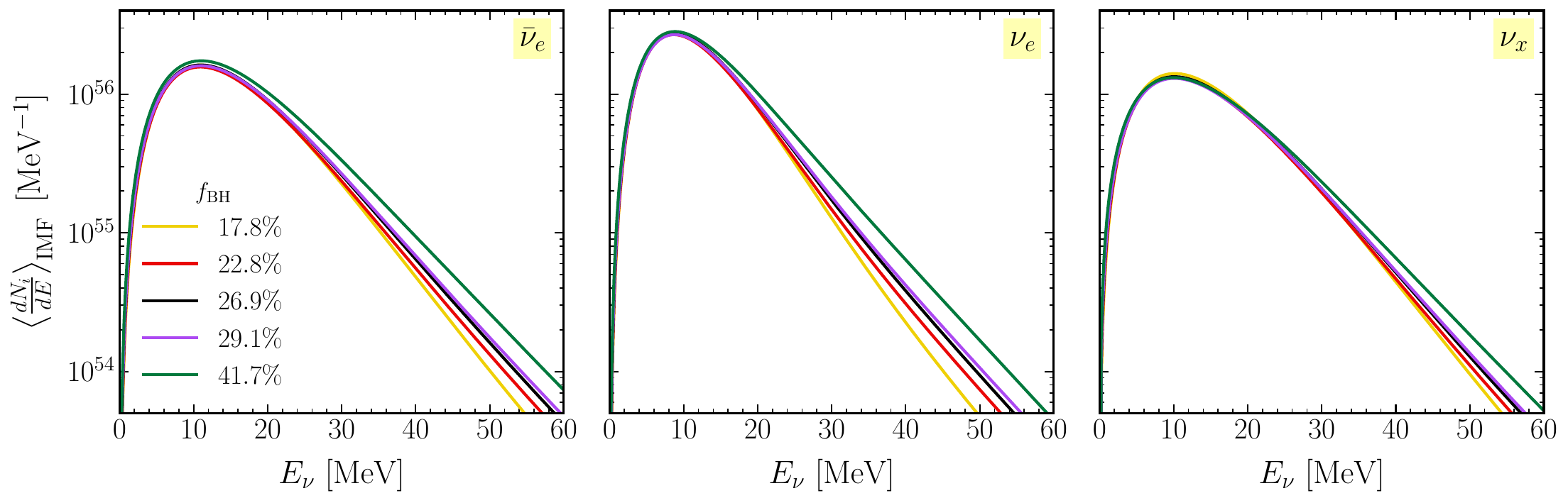}
    \caption{IMF-weighted neutrino emission spectra for $\bar{\nu}_e$ (left), $\nu_e$ (center), and $\nu_x$ (right), adapted from ref.~\cite{Kresse:2020nto}. The colored curves denote different black-hole formation fractions: $f_{\rm BH} = 17.8\%$ (yellow), $22.8\%$ (red), $26.9\%$ (black/fiducial), $29.1\%$ (purple), and $41.7\%$ (green).}
    \label{fig:nuspectra}
\end{figure}
For our DSNB flux calculations, we adopt the IMF-weighted neutrino spectra from the Garching group's core-collapse simulations, which include progenitors that lead to both neutron-star and black-hole formation.\footnote{The data are available via their supernova archive \url{https://wwwmpa.mpa-garching.mpg.de/ccsnarchive}.} Their model set covers ZAMS masses of $8.7$--$120M_\odot$ and black hole formation fractions $f_{\rm BH}=17.8\%$--$41.7\%$ (see ref.~\cite{Kresse:2020nto} for a detailed description). Figure~\ref{fig:nuspectra} displays, from left to right, the resulting IMF-weighted spectra for $\bar\nu_e$, $\nu_e$, and $\nu_x$. The different curves correspond to assumed black hole formation fractions $f_{\rm BH}=17.8\%,\,22.8\%,\,26.9\%,\,29.1\%,$ and $41.7\%$, shown in yellow, red, black, purple, and green, respectively. Hereafter, we adopt their fiducial model, which corresponds to $f_\text{BH}=26.9\%$. 
 
\paragraph{Flavor Conversion in the Stellar Envelope}~\\
\noindent The neutrino spectra arising from the core of the star are altered by the Mikheyev-Smirnov-Wolfenstein (MSW) flavor conversion in the stellar envelope due to coherent neutrino-matter interactions~\cite{Wolfenstein:1977ue,Mikheyev:1985zog,Dighe:1999bi}. Assuming adiabatic propagation through a smooth density profile, neutrinos emerging from the star are mapped onto the vacuum mass eigenstates at the stellar surface. The resulting mass-basis source spectra depend on the neutrino mass ordering. We follow the standard level-crossing scheme of ref.~\cite{Dighe:1999bi} and assume identical spectra for $\nu_\mu$ and $\nu_\tau$ (and likewise for antineutrinos), denoted collectively by $\nu_x$. For normal mass ordering (NO), the emitted flavor spectra are mapped onto the mass eigenstates as
\begin{align}\label{eq:spectraNO}
    \frac{d N_{\nu_1}}{dE}=\left\langle\frac{dN_{\nu_x}}{dE}\right\rangle_{\rm IMF},\quad
    \frac{d N_{\nu_2}}{dE}=\left\langle\frac{dN_{\nu_x}}{dE}\right\rangle_{\rm IMF},\quad
    \frac{d N_{\nu_3}}{dE}&=\left\langle\frac{dN_{\nu_e}}{dE}\right\rangle_{\rm IMF},\nonumber\\
    \frac{d N_{\bar\nu_1}}{dE}=\left\langle\frac{dN_{\bar\nu_e}}{dE}\right\rangle_{\rm IMF},\quad
    \frac{d N_{\bar\nu_2}}{dE}=\left\langle\frac{dN_{\nu_x}}{dE}\right\rangle_{\rm IMF},\quad
    \frac{d N_{\bar\nu_3}}{dE}&=\left\langle\frac{dN_{\nu_x}}{dE}\right\rangle_{\rm IMF},
\end{align}
while for inverted mass ordering (IO), the corresponding mapping can be written as
\begin{align}\label{eq:spectraIO}
    \frac{d N_{\nu_1}}{dE}=\left\langle\frac{dN_{\nu_x}}{dE}\right\rangle_{\rm IMF},\quad
    \frac{d N_{\nu_2}}{dE}=\left\langle\frac{dN_{\nu_e}}{dE}\right\rangle_{\rm IMF},\quad
    \frac{d N_{\nu_3}}{dE}&=\left\langle\frac{dN_{\nu_x}}{dE}\right\rangle_{\rm IMF},\nonumber\\
    \frac{d N_{\bar\nu_1}}{dE}=\left\langle\frac{dN_{\nu_x}}{dE}\right\rangle_{\rm IMF},\quad
    \frac{d N_{\bar\nu_2}}{dE}=\left\langle\frac{dN_{\nu_x}}{dE}\right\rangle_{\rm IMF},\quad
    \frac{d N_{\bar\nu_3}}{dE}&=\left\langle\frac{dN_{\bar\nu_e}}{dE}\right\rangle_{\rm IMF}.
\end{align}
\noindent We neglect the collective neutrino oscillations, which arise from neutrino-neutrino forward scattering in the dense region near the proto-neutron star. In the DSNB context, these effects are highly model dependent. They are expected to modify the final flux at only $\sim 5$--$10\%$ level, with their impact further diluted by the averaging over progenitors, emission phases, and source redshifts inherent in the DSNB~\cite{Chakraborty:2008zp,Lunardini:2012ne}.

\subsection{Cosmic Core-collapse Rate}
The core-collapse rate density can be inferred from the cosmic star formation rate density (SFRD), $\dot\rho_\star(z)$, which is the total stellar mass formed per unit time per unit comoving volume at redshift $z$. Observationally, the SFRD is reconstructed from multiwavelength galaxy surveys, typically combining UV, optical, and infrared tracers with corrections for dust attenuation and an assumed IMF~\cite{Hopkins:2006bw,Madau:2014bja}.
Assuming that all stars above the minimum progenitor mass undergo core collapse, the core-collapse rate is related to the SFRD by
\begin{align}\label{eq:ccsn}
    R_\text{CC}(z)=\frac{\int_{8.7M_\odot}^{125M_\odot}\psi(M)dM}{\int_{0.1M_\odot}^{125M_\odot}M\psi(M)dM}\,\dot\rho_\star(z)\simeq\frac{\dot\rho_\star(z)}{116M_\odot}.
\end{align}
where the IMF $\psi(M)$ is given by eq.~(\ref{eq:imf}) and the integration limits are chosen to match the progenitor mass range used for the neutrino-emission spectra. The numerical prefactor depends only weakly on the IMF, since the IMF dependence largely cancels between the SFRD calibration and the core-collapse rate conversion, provided both adopt the same IMF~\cite{Ekanger:2023qzw}. For the redshift evolution of the SFRD, we adopt the smooth broken power-law fit~\cite{Yuksel:2008cu,Horiuchi:2008jz,Mathews:2014qba}
\begin{align}\label{eq:sfr}
\dot\rho_\star(z)=\dot\rho_0\left[(1+z)^{\alpha\eta}+\left(\frac{1+z}{B}\right)^{\beta\eta}+\left(\frac{1+z}{C}\right)^{\gamma\eta}\right]^{1/\eta},
\end{align}
where $\dot\rho_0$ is the normalization factor in units of $M_\odot\, \text{yr}^{-1}\text{Mpc}^{-3}$. The parameters $B=(1+z_1)^{1-\alpha/\beta}$ and $C=(1+z_1)^{(\beta-\alpha)/\gamma}(1+z_2)^{1-\beta/\gamma}$ encode the transitions at the break redshifts $z_1\simeq1$ and $z_2\simeq4$, which mark the peak and subsequent decline of the SFRD. The parameter $\eta<0$ controls the smoothness of the transitions. For the Salpeter-A IMF, the parameter ranges listed in table~\ref{tab:sfr} are chosen to construct a conservative envelope that accounts for the scatter in the observational data (see ref.~\cite{Mathews:2014qba} for details).
 
\begin{table}[!h]
\vspace{2mm}
\caption{Parameters of the smooth broken power law SFRD fit of eq.~(\ref{eq:sfr}), taken from ref.~\cite{Mathews:2014qba}.}
\centering
\setlength{\tabcolsep}{7pt}
\begin{tabular}{lccccccc}
\toprule
 & $\dot{\rho}_0$ & $\eta$ & $\alpha$ & $\beta$ & $\gamma$ & $B$ & $C$ \\
\midrule
Upper limit & 0.0200 & -10 & 4.22 & 0.14  & -9.0   & $1.25\times10^{-8}$ & 6.87 \\[0.1em]
Best Fit   & 0.0104 & -10 & 4.22 & -0.20 & -11.30 & $2.70\times10^{6}$  & 6.37 \\[0.1em]
Lower limit & 0.0069 & -10 & 4.22 & -0.42 & -13.60 & $1.6\times10^{3}$   & 5.90 \\
\bottomrule
\end{tabular}
\label{tab:sfr}
\end{table}

\subsection{Neutrino Propagation and the DSNB Flux}
To compute the DSNB spectrum at Earth, we evolve the phase space distribution function $f_i(p,t)$ for each mass eigenstate $\nu_i$, where $p=|\vec{p}|$ is the physical momentum magnitude and $t$ is cosmic time. Homogeneity and isotropy of the DSNB allow $f_i$ to depend only on $p$ and $t$. The differential number density, defined as the number of neutrinos per comoving volume per unit momentum, is related to the phase space distribution function by~\cite{Ando:2023fcc}
\begin{align}\label{eq:ni}
    n_i(p,t)=a^3(t)\,\frac{p^2}{2\pi^2}\,f_i(p,t)\,,
\end{align}
where $a(t)$ is the cosmological scale factor. For relativistic neutrinos ($E\simeq p$), the all-sky differential flux observed today is simply the present-day number density
\begin{align}\label{eq:flux_n}
    \Phi_i(E)=c\,n_i(p,t_0)\,,
\end{align}
with $t_0$ the present time. The evolution of $f_i$ is governed by the Boltzmann equation. In a homogeneous and isotropic Friedmann--Robertson--Walker universe, it can be written as 
\begin{align}\label{eq:BE0}
    \left(\partial_t-H(t)\,p\,\partial_p\right)f_i(p,t)=\mathcal{C}_i(p,t)\,,
\end{align}
where $H=\dot{a}/a$ is the Hubble parameter and $\mathcal{C}_i$ denotes the collision term. In the standard free-streaming case ($\nu$SI absent), the only source is core-collapse emission,
\begin{align}\label{eq:source}
    \mathcal{C}_i(p,t)=\mathcal{S}_{\nu_i}(p,t)
    =\frac{2\pi^2}{p^2\,a^3(t)}\,R_{\text{CC}}(z)\,
    \frac{dN_{\nu_i}}{dE}\bigg|_{E=p}\,,
\end{align}
where $R_{\rm CC}(z)$ is the core-collapse rate of eq.~(\ref{eq:ccsn}) and $dN_{\nu_i}/dE$ denotes the IMF-averaged, MSW converted mass basis emission spectra of eqs.~(\ref{eq:spectraNO}) and (\ref{eq:spectraIO}), evaluated at the physical momentum $p$ of the neutrino at the time of emission.

Eq.~(\ref{eq:BE0}) can be solved analytically by transforming to comoving variables. Defining the comoving momentum $\tilde{p}\equiv p\,a(t)=p/(1+z)$ and using the cosmological redshift $z=1/a-1$ as the time variable, the partial derivatives transform as
\begin{align}\label{eq:optrans}
    \partial_p\to(1+z)^{-1}\,\partial_{\tilde{p}}\,,\qquad
    \partial_t\to H\,\tilde{p}\,\partial_{\tilde{p}}-(1+z)\,H\,\partial_z\,.
\end{align}
Substituting into eq.~(\ref{eq:BE0}), the Hubble drag term $-H\,p\,\partial_p f_i$ cancels the $H\,\tilde{p}\,\partial_{\tilde{p}}f_i$ contribution from $\partial_tf_i$, and the Boltzmann equation reduces to
\begin{align}\label{eq:BEz}
    -H(z)\,\partial_z f_i(\tilde{p},z) =\frac{2\pi^2}{\tilde{p}^2}\,R_{\text{CC}}(z)\, \frac{dN_{\nu_i}}{dE}\bigg|_{E=\tilde{p}(1+z)}\,.
\end{align}
Integrating it from the present epoch ($z=0$) to arbitrarily high redshift, with the boundary condition $f_i\to0$ as $z\to\infty$, gives the present-day phase-space distribution
\begin{align}\label{eq:fsol}
    f_i(\tilde{p},z=0)=\frac{2\pi^2}{\tilde{p}^2}\int_0^{\infty}\frac{dz}{H(z)}\, R_{\text{CC}}(z)\,
    \frac{dN_{\nu_i}}{dE}\bigg|_{E=\tilde{p}(1+z)}\,.
\end{align}
At $z=0$, the comoving momentum coincides with the observed momentum, $\tilde{p}=p_0=E_0$, where $E_0$ is the neutrino energy measured at Earth. Substituting eq.~(\ref{eq:fsol}) into Eqs.~(\ref{eq:ni}) and (\ref{eq:flux_n}) and using the Friedmann equation $H(z)=H_0\sqrt{\Omega_m(1+z)^3+\Omega_\Lambda}$, the DSNB flux for mass eigenstate $\nu_i$ is
\begin{align}\label{eq:stddsnb}
    \Phi_i(E_0)=\frac{c}{H_0}\int_0^{\infty}\frac{dz}{\sqrt{\Omega_m(1+z)^3+\Omega_\Lambda}}\, R_{\text{CC}}(z)\,\frac{dN_{\nu_i}}{dE}\bigg|_{E=E_0(1+z)}\,,
\end{align}
where $H_0\approx67.4\text{ km } \text{s}^{-1}\text{ Mpc}^{-1}$ is the present-day value of the Hubble parameter, and $\Omega_m=0.315$, $\Omega_\Lambda=0.685$ are the matter and dark energy density fractions, respectively~\cite{Planck:2018vyg}. 
We have neglected the radiation contribution, which is negligible over the redshift range relevant for the DSNB.

Although eq.~(\ref{eq:stddsnb}) provides the DSNB flux in the mass basis, neutrinos are experimentally detected in the flavor basis. Since the mass eigenstates propagate incoherently over cosmological distances, the flavor-basis flux is obtained by the oscillation-averaged projection
\begin{align}\label{eq:flavproj}
    \Phi_{\nu_\alpha}(E_0)=\sum_i|U_{\alpha i}|^2\,\Phi_i(E_0)\,,
\end{align}
where $U$ is the Pontecorvo--Maki--Nakagawa--Sakata (PMNS) neutrino mixing matrix. Throughout this analysis, the neutrino oscillation parameters entering the PMNS matrix are fixed to $\theta_{12}=33.4^{\circ}$, $\theta_{13}=8.57^{\circ}$, $\theta_{23}=49.2^{\circ}$, and $\delta_{\rm CP}=195^{\circ}$ which lie within the $3\sigma$ ranges reported in NuFit~\cite{Esteban:2020cvm,Esteban:2024eli}.\footnote{The Majorana phases are set to zero throughout this work.}

The resulting free-streaming fluxes are shown as solid lines in figure~\ref{fig:dsnb_std} for all six flavor states, for both NO (red) and IO (blue).
The difference between NO and IO is most pronounced in the $\nu_e$ and $\bar\nu_e$ channels: in the neutrino sector, the MSW conversion maps $\nu_e$ onto $\nu_3$ in NO and onto $\nu_2$ in IO, while in the antineutrino sector it maps $\bar\nu_e$ onto $\bar\nu_1$ in NO and onto $\bar\nu_3$ in IO. In contrast, the non-electron flavors show only a mild ordering dependence. A notable feature is the splitting between $\Phi_{\nu_\mu}$ and $\Phi_{\nu_\tau}$ (and likewise between $\Phi_{\bar\nu_\mu}$ and $\Phi_{\bar\nu_\tau}$), visible in figure~\ref{fig:dsnb_std}. This arises from the deviation of $\theta_{23}$ from maximality together with non-zero $\theta_{13}$, which in turn break the $\mu$--$\tau$ symmetry of the PMNS projection. This effect is absent in analyses that employ a two-flavor approximation or assume $\theta_{23}=45^\circ$.

This free-streaming solution serves as the baseline for the next section, where $\nu$SI introduces additional collision terms in eq.~(\ref{eq:BE0}) that modify the propagation of the DSNB mass eigenstates during their transit through the C$\nu$B.

\section{\label{sec:nusi}Secret Neutrino Interactions and the DSNB}
In this work, we study non-standard neutrino self-interactions using an effective Lagrangian framework to ensure a model-independent treatment. We assume that neutrinos are Majorana particles, with masses arising from an underlying mechanism, e.g., a seesaw scenario. For simplicity, we consider the light $\nu$SI mediator $\phi$ to be a real scalar with mass $m_\phi$. Although vector mediators have also been explored in literature~\cite{Bakhti:2023mvo,Kelly:2018tyg,Laha:2013xua}, they are typically subject to more stringent constraints.  We therefore restrict our focus to the scalar case. The effective interaction Lagrangian in the flavor basis can be written as,
\begin{align}
  \mathcal{L}\supset-\frac12m^2_\phi\phi^2+\frac12\sum_{\alpha,\beta}\left(g_{\alpha\beta}\,\phi\,\nu_\alpha\nu_\beta+{\rm h.c.}\right)\,,
\end{align}
where $\nu_\alpha$ denotes a two-component Weyl spinor and $g_{\alpha\beta}=g_{\beta\alpha}$ is the symmetric $\nu$SI coupling matrix in the flavor basis ($\alpha,\beta=e,\mu,\tau$). Although $g_{\alpha\beta}$ is in principle a general $3\times3$ symmetric matrix, we restrict our analysis to flavor diagonal couplings ($g_{\alpha\beta}=0$ for $\alpha\neq\beta$). Specifically, we consider four representative coupling structures: 
\begin{enumerate}
    \item Universal couplings: $g_{\alpha\beta}=g_{\rm univ}\,{\rm diag}\,(1,1,1)$,
    \item $e$-specific couplings: $g_{\alpha\beta}={\rm diag}\,(g_{ee},0,0)$,
    \item $\mu$-specific couplings: $g_{\alpha\beta}={\rm diag}\,(0,g_{\mu\mu},0)$,
    \item $\tau$-specific couplings: $g_{\alpha\beta}={\rm diag}\,(0,0,g_{\tau\tau})$.
\end{enumerate}
\par As discussed in section~\ref{sec:dsnb}, the neutrinos emitted from the supernova core undergo MSW conversion in the stellar envelope and exit the star as incoherent mass eigenstates. The scattering of DSNB neutrinos off the C$\nu$B is therefore most naturally formulated in the mass basis. Using the transformation relation between flavor and mass eigenstates, $\nu_\alpha=\sum_{i=1}^{3} U_{\alpha i}\,\nu_i$, the coupling structures in the mass basis ($g_{ij}$) can be written as
\begin{align}\label{eq:coup}
g_{ij} = \sum_{\alpha,\beta} g_{\alpha\beta}\,
U_{\alpha i} U_{\beta j} \, ,
\end{align}
where $\nu_i~(i,j=1,2,3)$ represent the three neutrino mass eigenstates. For the diagonal flavor structures considered here, the above expression simplifies to
\begin{align}\label{eq:coup_diag}
    g_{ij}=\sum_{\alpha}g_{\alpha\alpha}\,U_{\alpha i}\,U_{\alpha j}\,.
\end{align}
Notice that even though we assume a purely diagonal structure for $g_{\alpha\beta}$ in the flavor basis, eq.~(\ref{eq:coup_diag}) implies that the mass-basis coupling matrix $g_{ij}$ is in general non-diagonal. 

The four coupling structures above are chosen to capture the qualitatively distinct depletion patterns accessible to flavor-diagonal scalar couplings. Since the resonant process $\nu_i\nu_k\to\phi$ depends on the mass-basis coupling $g_{ik}$, each structure assigns different weights to the PMNS columns, leading to distinct, ordering-dependent imprints in the DSNB flux. For the universal scenario, $g_{ij}=g_{\rm univ}\sum_{\alpha}U_{\alpha i}U_{\alpha j}$, which is generally non-diagonal in presence of non-zero phase because $\sum_{\alpha}U_{\alpha i}U_{\alpha j}\neq\delta_{ij}$ for a complex mixing matrix. However, the off-diagonal entries are suppressed by $\sin\theta_{13}$, giving rise to an effectively diagonal coupling. This structure arises in $U(1)_{B-L}$-like extensions and serves as a natural benchmark. The flavor-specific cases, corresponding to gauged $U(1)_{L_\alpha}$ scenarios, yield non-diagonal mass-basis matrices weighted by the PMNS row $U_{\alpha i}$, enabling inter-eigenstate transitions that are strongly suppressed in the universal case.

The free-streaming DSNB flux derived in section~\ref{sec:dsnb} is modified by $\nu$SI-mediated resonant scattering of DSNB neutrinos on the C$\nu$B, $\nu_i\nu_k\to\phi\to\nu_j\nu_l$. The resulting spectral distortion depends critically on whether the target C$\nu$B species is relativistic or non-relativistic today. For a non-relativistic C$\nu$B target, the resonance condition, $E_{\rm res}\simeq m_\phi^2/(2m_\nu)$, produces a sharp dip at a single DSNB energy. In the relativistic regime, by contrast, the center-of-mass energy
\begin{align}\label{eq:res}
    s = 2E_\nu E_{\rm C\nu B}\,(1-\cos\theta)
\end{align}
depends on both the C$\nu$B energy $E_{\rm C\nu B}$, distributed thermally according to the Fermi--Dirac spectrum at temperature $T_{\rm C\nu B}^0\approx1.95~\mathrm{K}\approx1.7\times10^{-4}~\mathrm{eV}$, and the scattering angle $\theta$. The resonance condition $s=m_\phi^2$ is then satisfied over a broad range of DSNB energies $E_\nu$, producing a wide absorption feature in the spectrum rather than a narrow dip.

The absolute neutrino mass scale remains unknown: direct kinematic measurements constrain the effective electron neutrino mass to $m_{\bar\nu_e}<0.45~\mathrm{eV}$~\cite{KATRIN:2024cdt}, while cosmological observations bound $\sum m_\nu\leq0.12~\mathrm{eV}$~\cite{Planck:2018vyg}. 
The lightest neutrino mass eigenstate may therefore be sufficiently light to remain relativistic today, with a broad thermal momentum distribution. We adopt this assumption throughout, as it leads to broader absorption features that significantly enhance DSNB sensitivity to $\nu$SI. In the following, we denote the lightest state by $\nu_k$ with $k=1$ for NO and $k=3$ for IO.

\subsection{Modified Boltzmann Equation}
To account for the distinct emission spectra of neutrinos and antineutrinos (eqs.~(\ref{eq:spectraNO})--(\ref{eq:spectraIO})), we track the phase space distribution function of each mass eigenstate separately. In the presence of $\nu$SI, the Boltzmann equation~(\ref{eq:BE0}) acquires additional collision terms describing the depletion and regeneration of DSNB neutrinos through their interaction with the C$\nu$B, as well as the production and decay of the mediator $\phi$. To simplify the notation, we collect all seven species---three neutrino mass eigenstates, three antineutrino mass eigenstates, and the scalar $\phi$---into a single distribution function vector as
\begin{align}
    \mathbf{F}(p,t)\equiv\big(f_{\nu_1},\,f_{\nu_2},\,f_{\nu_3},\,f_{\bar\nu_1},\,f_{\bar\nu_2},\,f_{\bar\nu_3},\,f_\phi\big)^T,
\end{align}
where each component depends on momentum $p$ and cosmic time $t$. Note that the C$\nu$B neutrinos are not included in $\mathbf{F}$; their energies are negligible compared to DSNB energies, so the C$\nu$B is treated as a static thermal background. The coupled system of Boltzmann equations then takes the following form
\begin{align}\label{eq:mod_BE}
    \left(\partial_t - Hp\,\partial_p\right)\mathbf{F}(p,t)=-\boldsymbol{\Gamma}^-(p,t)\,\mathbf{F}(p,t)+\boldsymbol{\Gamma}^+(p,t)+\boldsymbol{\mathcal{S}}(p,t)\,,
\end{align}
where the source, creation-rate, and depletion-rate operators are given as
\begin{align}
\boldsymbol{\mathcal{S}} &= \big(\mathcal{S}_{\nu_1},\,\mathcal{S}_{\nu_2},\,\mathcal{S}_{\nu_3},\,\mathcal{S}_{\bar\nu_1},\,\mathcal{S}_{\bar\nu_2},\,\mathcal{S}_{\bar\nu_3},\,0\big)^T, \\
\boldsymbol{\Gamma}^+ &= \big(\Gamma^+_{\nu_1},\,\Gamma^+_{\nu_2},\,\Gamma^+_{\nu_3},\,\Gamma^+_{\bar\nu_1},\,\Gamma^+_{\bar\nu_2},\,\Gamma^+_{\bar\nu_3},\,\Gamma^+_\phi\big)^T, \\
\boldsymbol{\Gamma}^- &= \mathrm{diag}\big(\Gamma^-_{\nu_1},\,\Gamma^-_{\nu_2},\,\Gamma^-_{\nu_3},\,\Gamma^-_{\bar\nu_1},\,\Gamma^-_{\bar\nu_2},\,\Gamma^-_{\bar\nu_3},\,\Gamma^-_\phi\big),
\end{align}
with $\mathcal{S}_{\nu_i}$ defined in eq.~\eqref{eq:source}.

In principle, all $\nu$SI-mediated processes contribute to these rates. However, under the narrow-width approximation, the dominant contribution comes from the resonant $s$-channel, which factorizes into the fusion $\nu_i\nu_k\to\phi$ and the subsequent decay $\phi\to\nu_j\nu_l$. This channel scales as $g^2$ in the cross section, $\sigma_{\rm res}\propto g^2\delta(s-m_\phi^2)$, whereas non-resonant processes such as $\nu\nu\to\nu\nu$ (t-channel) and $\nu\nu\to\phi\phi$ scale as $g^4$ and can be safely neglected in the weak-coupling regime of interest.

The individual rates appearing in eq.~(\ref{eq:mod_BE}) were derived in ref.~\cite{Wang:2025qap}. The production rate of $\nu_i$ (and $\bar\nu_i$) from $\phi$ decay is
\begin{align}\label{eq:Gp_nu}
    \Gamma^+_{\nu_i}=\Gamma^+_{\bar\nu_i}=\frac{1}{2}\,\frac{\sum_{j}|g_{ij}|^2}{16\pi}\,\frac{2m_\phi^2}{E_\nu^2}\int_{E_\phi^{\rm min}}^{\infty}dE_\phi\,f_\phi\,,
\end{align}
where the equality $\Gamma^+_{\nu_i}=\Gamma^+_{\bar\nu_i}$ follows from the Majorana nature of the neutrinos and the Hermiticity of the interaction. The production rate of $\phi$ from the fusion $\nu_i\nu_k\to\phi$ (and the conjugate process) is
\begin{align}\label{eq:Gp_phi}
    \Gamma^+_\phi=\sum_i|g_{ik}|^2\,\frac{1}{16\pi}\,\frac{m_\phi^2}{E_\phi p_\phi}\int_{E_\nu^-}^{E_\nu^+}dE_\nu\,\big(f_{\nu_i}+f_{\bar\nu_i}\big)\,\exp\left[-\frac{E_\phi-p_\nu}{T_{\rm C\nu B}}\right],
\end{align}
where the sum runs over all DSNB mass eigenstates $\nu_i$ scattering off the lightest C$\nu$B state $\nu_k$. Here, $E_\nu$ and $p_\nu$ denote the energy and momentum of the DSNB neutrino, while $E_\phi$ and $p_\phi$ denote those of the scalar mediator. From the energy-momentum conservation, one obtains the kinematic limits $E_\phi^{\rm min}=E_\nu+m_\phi^2/(4E_\nu)$ for the minimum mediator energy, and $E_\nu^\pm=(E_\phi\pm p_\phi)/2$ for the integration bounds in eq.~(\ref{eq:Gp_phi}). The depletion rate of $\nu_i$ (and $\bar\nu_i$) due to the fusion $\nu_i\nu_k\to\phi$ is
\begin{align}\label{eq:Gm_nu}
    \Gamma^-_{\nu_i}=\Gamma^-_{\bar\nu_i}=\frac{|g_{ik}|^2}{16\pi}\,\frac{m_\phi^2\,T_{\rm C\nu B}}{E_\nu^2}\,\exp\left[-\frac{m_\phi^2}{4E_\nu T_{\rm C\nu B}}\right],
\end{align}
where the exponential factor reflects the thermal distribution of the C$\nu$B targets: for $E_\nu\gg m_\phi^2/(4T_{\rm C\nu B})$ the depletion rate saturates, while for lower energies it is exponentially suppressed, defining the characteristic energy scale of the absorption feature. The depletion rate of $\phi$ from its decay into all neutrino pairs is
\begin{align}\label{eq:Gm_phi}
    \Gamma^-_\phi=\frac{1}{16\pi}\,\frac{m_\phi^2}{E_\phi}\,\sum_{i,j}|g_{ij}|^2\,.
\end{align}
The C$\nu$B temperature scales with redshift as $T_{\rm C\nu B}(z)=T_{\rm C\nu B}^0\,(1+z)$, with $T_{\rm C\nu B}^0\approx1.95~\mathrm{K}$. The rates in eqs.~(\ref{eq:Gp_nu})–(\ref{eq:Gm_phi}) are all evaluated at the physical values of the momenta, energies, and temperature. 

The coupled Boltzmann equation is solved numerically following the discretized momentum grid procedure as described in ref.~\cite{Wang:2025qap}. The solution provides the present day phase-space distributions $f_{\nu_i}(p,t_0)$, from which the mass-basis fluxes $\Phi_i(E_0)$ are computed using eq.~\eqref{eq:flux_n}. The observed flavor-basis fluxes then follow from the oscillation-averaged projection of eq.~(\ref{eq:flavproj}). 

\subsection{\label{sec:mod_spectra}Spectral Features}
\begin{figure}[!b]
    \centering
    \includegraphics[width=\linewidth]{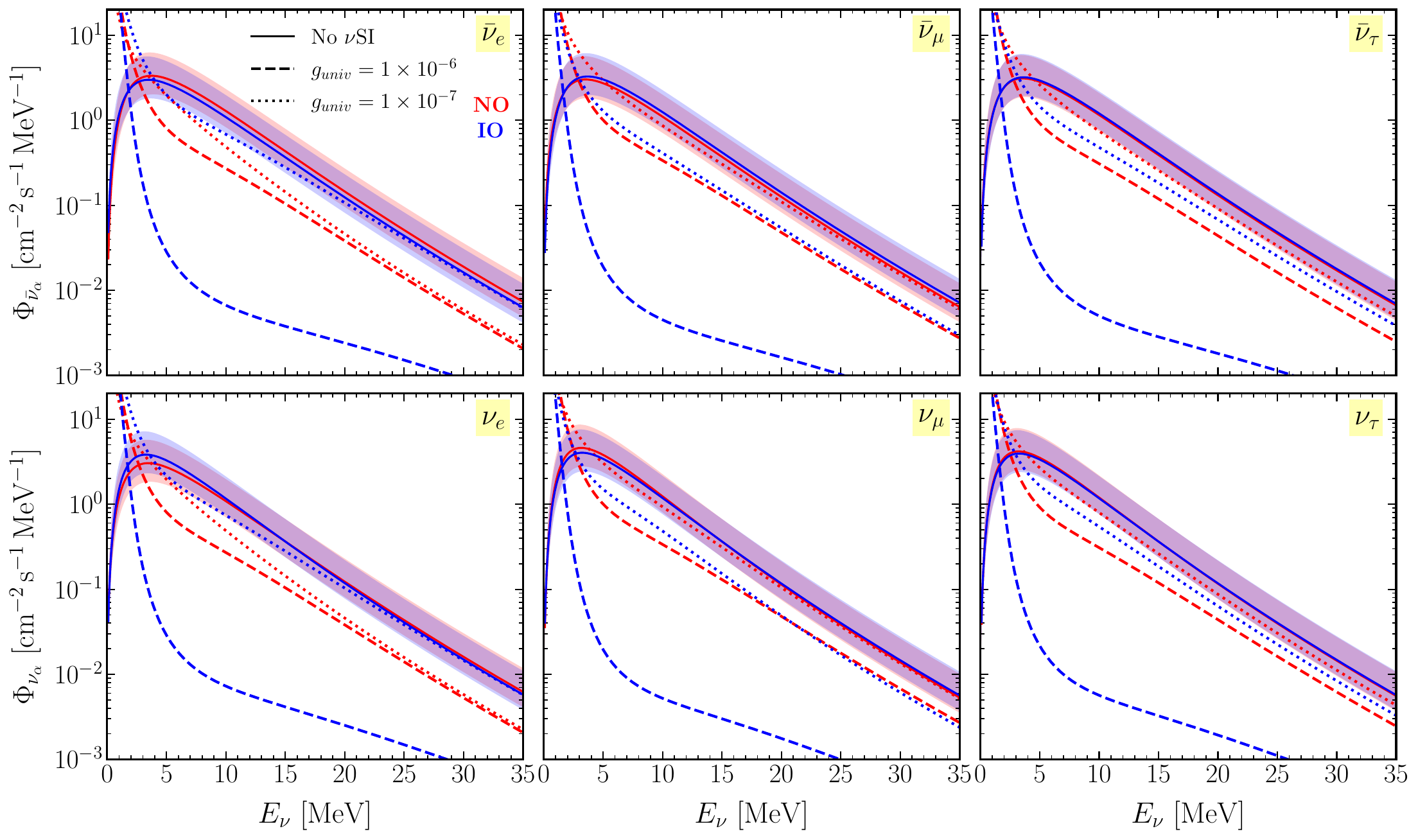}
    \caption{DSNB flux spectra $(\Phi_{\nu_{\alpha}})$ for all six flavor states as a function of neutrino energy, for both NO (red) and IO (blue). Solid lines show the free-streaming solution using the best-fit SFRD parameters of table~\ref{tab:sfr}, with the shaded bands indicating the upper and lower parameter ranges. The dashed and dotted lines show the $\nu$SI-modified spectra for a universal coupling with mediator mass $m_\phi$=100 eV and coupling $g_{\rm univ}=10^{-6}$ and $10^{-7}$, respectively, computed using the best-fit SFRD parameters only.}
    \label{fig:dsnb_std}
\end{figure}

\begin{figure}[!h]
    \centering
    \includegraphics[width=\linewidth]{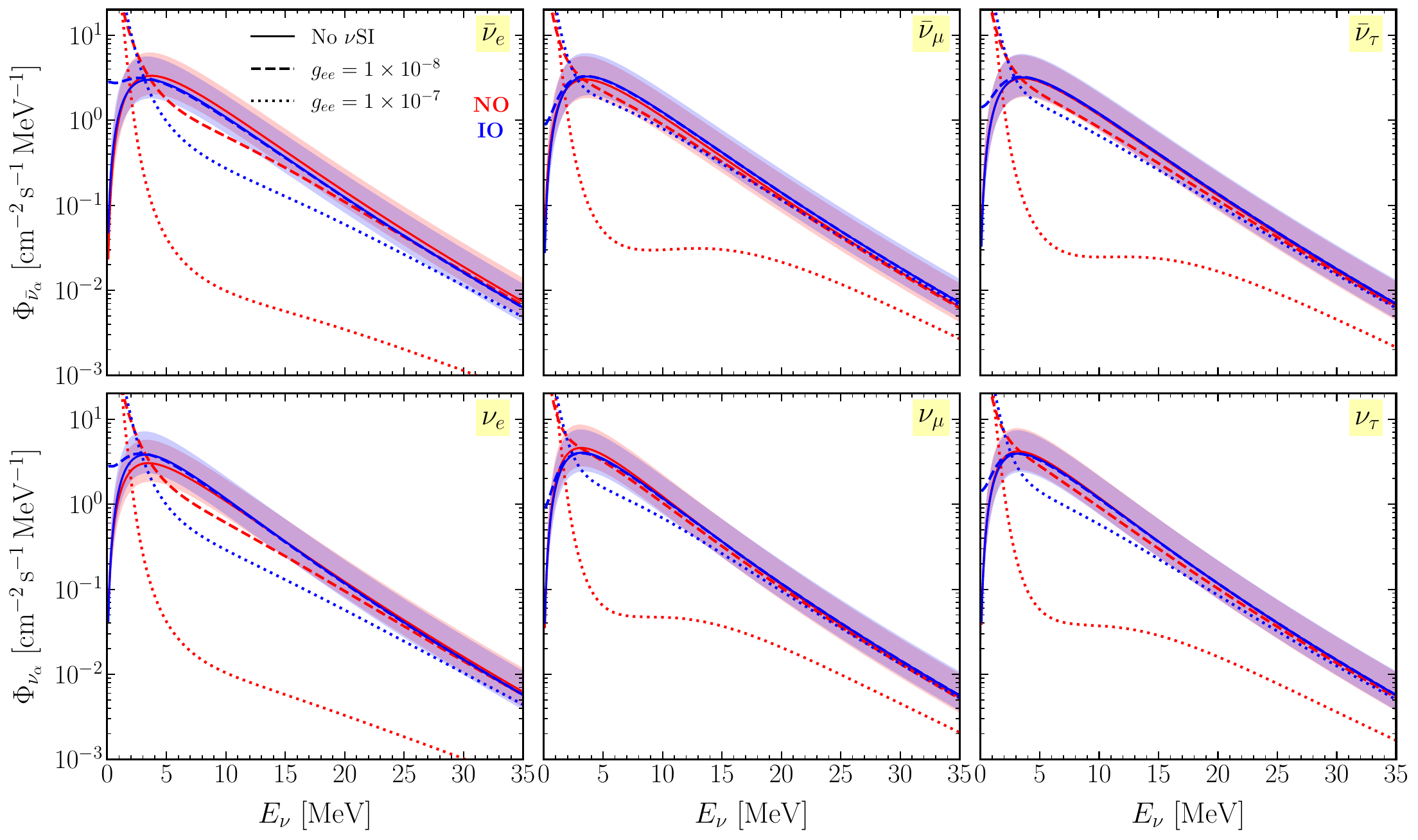}
    \caption{Same as figure~\ref{fig:dsnb_std} but for $e$-specific coupling $g_{\rm ee}=10^{-7}$ (dotted), $10^{-8}$ (dashed) with $m_\phi$=100 eV.}
    \label{fig:dsnb_std_gee}
\end{figure}

\begin{figure}[!h]
    \centering
    \includegraphics[width=\linewidth]{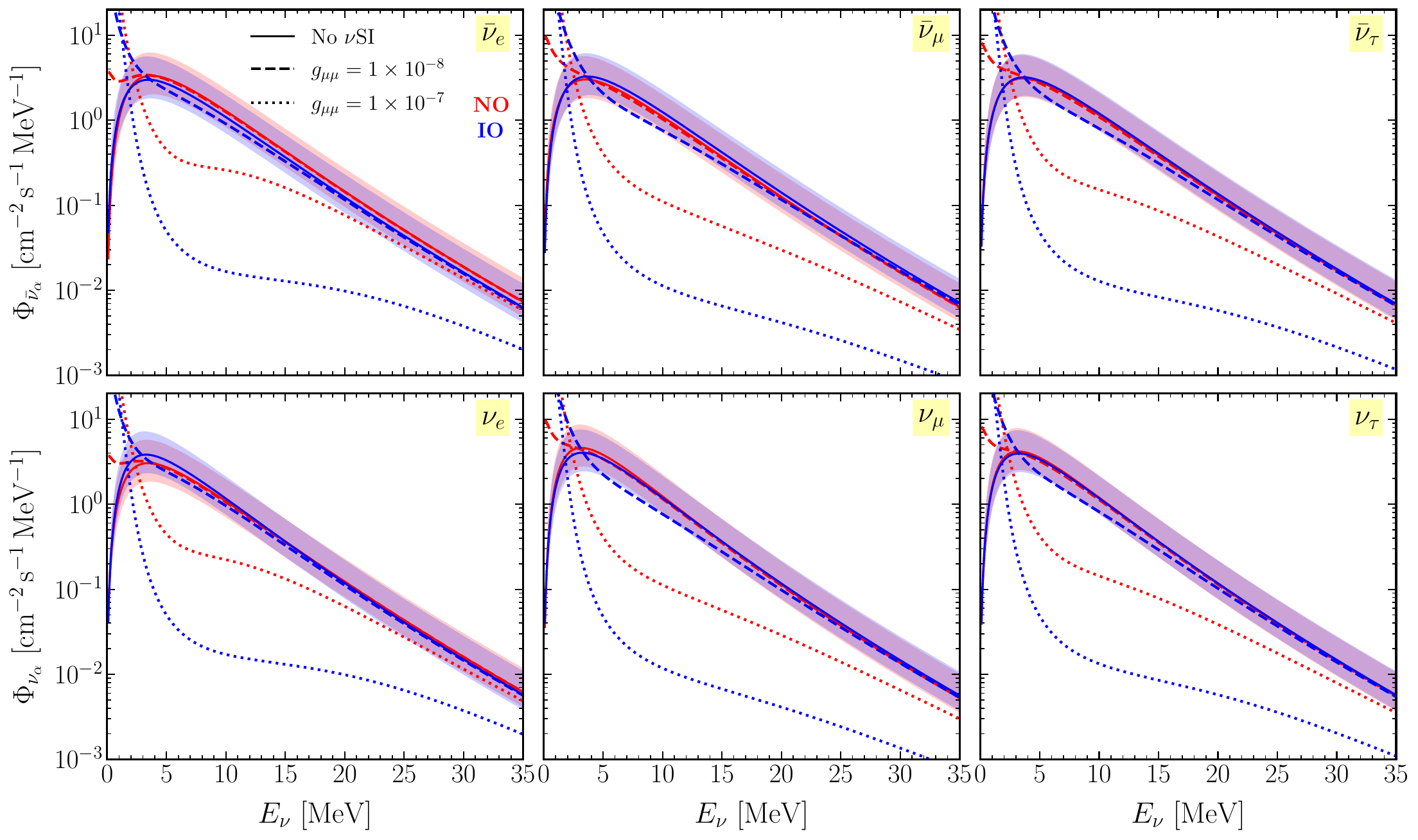}
    \caption{Same as figure~\ref{fig:dsnb_std} but for $\mu$-specific coupling $g_{\rm \mu\mu}=10^{-7}$ (dotted), $10^{-8}$ (dashed).}
    \label{fig:dsnb_std_gmm}
\end{figure}

\begin{figure}[!h]
    \centering
    \includegraphics[width=\linewidth]{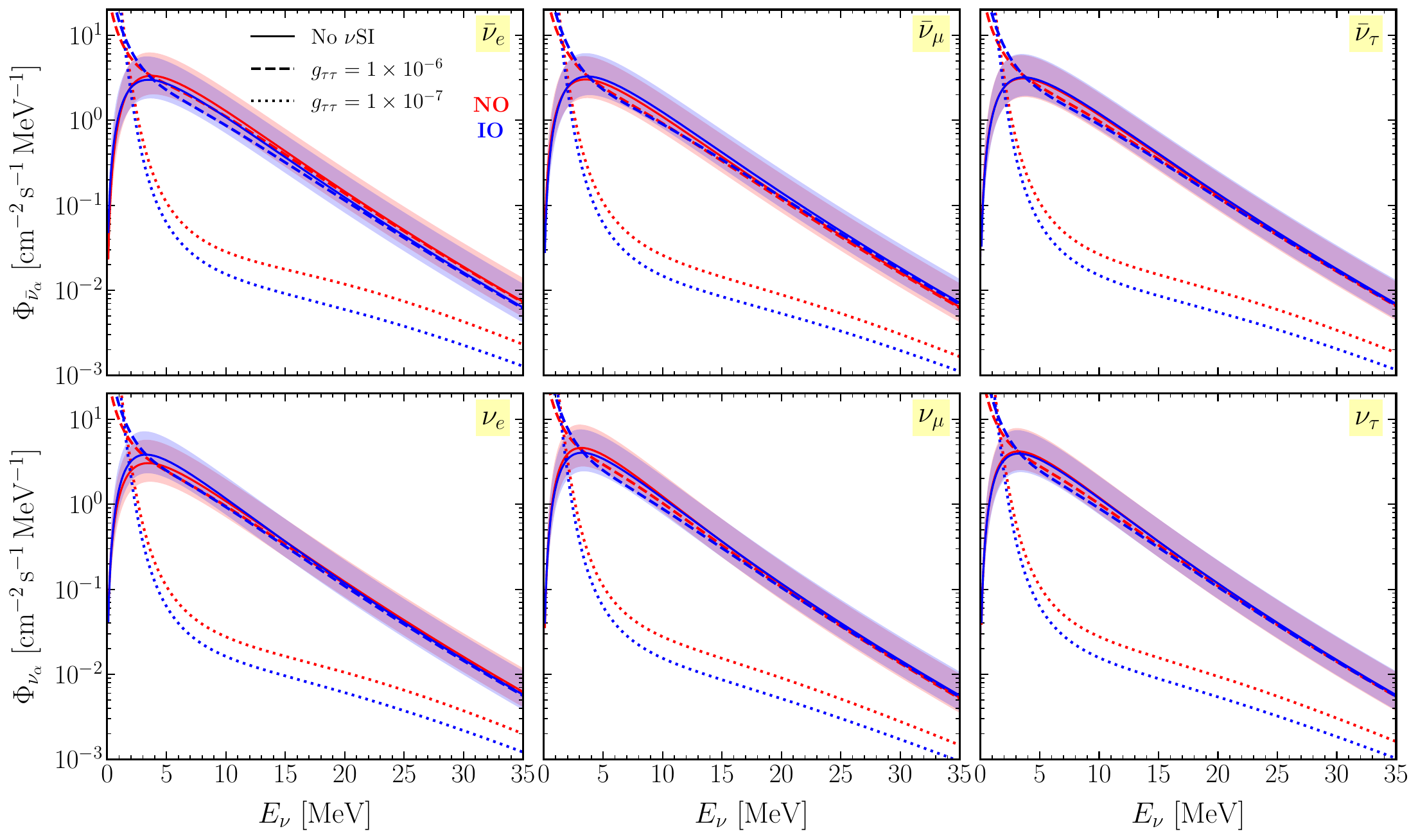}
    \caption{Same as figure~\ref{fig:dsnb_std} but for $\tau$-specific coupling $g_{\rm \tau\tau}=10^{-7}$ (dotted), $10^{-8}$ (dashed).}
    \label{fig:dsnb_std_gtt}
\end{figure}

In figures~\ref{fig:dsnb_std}--\ref{fig:dsnb_std_gtt}, we present the $\nu$SI-modified DSNB fluxes for all four coupling structures with $m_\phi=100$~eV and two benchmark coupling values, alongside the free-streaming scenario. The NO and IO scenarios are depicted by red and blue lines, respectively. In all cases, the resonant process $\nu_i\nu_k\to\phi$ depletes the DSNB flux over a broad range of energies, since the relativistic C$\nu$B target $\nu_k$ allows the resonance condition $s=m_\phi^2$ to be satisfied for a continuum of DSNB energies. The mediator subsequently decays, $\phi\to\nu_j\nu_l$, with branching fractions proportional to $|g_{jl}|^2/\sum_{mn}|g_{mn}|^2$, redistributing the absorbed energy among mass eigenstates at lower energies. The net observable effect is a spectral deficit at the resonance energies, accompanied by a pile-up below $\sim5$~MeV, well below the signal windows of JUNO, HK, and DUNE.

The detailed shape of the modified spectra emerges from the full solution of the coupled Boltzmann system in eq.~(\ref{eq:mod_BE}), which generates a rich interplay of depletion, redistribution among mass eigenstates, and projection onto the flavor states. At a qualitative level, however, the dominant features can be understood from two factors: (i)~the depletion rate of each mass eigenstate, $\Gamma^-_{\nu_i}\propto |g_{ik}|^2$, which depends on the coupling structure and the mass ordering through the identity of the lightest state $\nu_k$; and (ii)~the PMNS projection $|U_{\alpha i}|^2$ that maps mass-eigenstate fluxes onto the observed flavor states. We discuss each coupling structure below.
\paragraph{Universal coupling:} 
The mass-basis couplings $g_{ij}$ are effectively diagonal in this case, with the off-diagonal entries $g_{13}$ and $g_{23}$ suppressed by $s_{13}s_\delta$ (and vanishing identically for $\delta_{\rm CP}=0$) and $g_{12}$ further suppressed by $s^2_{13}$.~\footnote{Throughout, $s_{ij}\equiv\sin\theta_{ij}$, $c_{ij}\equiv\cos\theta_{ij}$, and $s_\delta\equiv\sin\delta_{\rm CP}$.} In NO ($k=1$), $\nu_1$ is therefore dominantly absorbed, with $s_{13}$-suppressed depletion of $\nu_3$ and doubly suppressed ($s_{13}^2$) depletion of $\nu_2$. In IO ($k=3$), $\nu_3$ is dominantly absorbed, while both $\nu_1$ and $\nu_2$ undergo $s_{13}$-suppressed depletion.

The ordering dependence of the observed flavor fluxes follows from the PMNS projection. For $\nu_\mu$ and $\nu_\tau$, IO is more strongly suppressed than NO across all coupling values as these flavors draw heavily from $\Phi_3$---which is the dominantly depleted state in IO---with weights $|U_{\mu 3}|^2\approx0.56$ and $|U_{\tau 3}|^2\approx0.42$. In contrast, they receive only small contributions from $\Phi_1$---the dominantly depleted state in NO---with $|U_{\mu 1}|^2\approx0.07$ and $|U_{\tau 1}|^2\approx0.25$.

For $\nu_e$, the pattern is coupling-dependent. At moderate couplings ($g_{\rm univ}\lesssim10^{-7}$), the dominant effect is the leading-order depletion of $\Phi_1$ in NO, which projects onto $\nu_e$ with the large weight $|U_{e1}|^2\approx0.68$; in IO, the dominant depletion of $\Phi_3$ is projected with the small weight $|U_{e3}|^2\approx0.022$, leaving $\nu_e$ relatively unaffected. At larger couplings ($g_{\rm univ}\gtrsim10^{-6}$), however, the sub-leading $s_{13}$-suppressed depletion of $\Phi_1$ and $\Phi_2$ accumulate substantially in IO, while $\Phi_2$ remains protected in NO by its $s_{13}^2$-suppressed coupling to $\nu_1$. Since all flavor fluxes receive significant contributions from $\Phi_2$ ($|U_{e2}|^2\approx0.30$, $|U_{\mu 2}|^2\approx0.37$, $|U_{\tau 2}|^2\approx0.34$), this ordering-dependent protection of $\Phi_2$ causes IO to become more depleted than NO even in the $\nu_e$ channel (see figure~\ref{fig:dsnb_std}). This coupling-dependent crossover in the NO/IO hierarchy in the $\nu_e$ channel has direct consequences for the projected sensitivities, as discussed in section~\ref{res}. 

The antineutrino channel follows an identical pattern, since the PMNS weights $|U_{\alpha i}|^2$ entering the flavor projection are the same for neutrinos and antineutrinos.
\paragraph{Flavor-specific couplings:} 
For a single non-zero diagonal entry $g_{\alpha\alpha}$, the mass-basis coupling reduces to $g_{ij}=g_{\alpha\alpha}\,U_{\alpha i}\,U_{\alpha j}$ and the depletion rate simplifies to $\Gamma^-_{\nu_i}\propto g_{\alpha\alpha}^2\,|U_{\alpha i}|^2\,|U_{\alpha k}|^2$. The overall interaction strength is thus governed by $|U_{\alpha k}|^2$, which depends on the mass ordering. Moreover, unlike the universal case, the off-diagonal entries of $g_{ij}$ are not suppressed, so \emph{all} mass eigenstates undergo appreciable resonant scattering with $\nu_k$.

\begin{enumerate}
    \item \textit{Electron-specific coupling}~($g_{ee}$):   
    In NO, $|U_{e1}|^2\approx0.68$ enhances all scattering rates and all flavor fluxes are visibly depleted. In IO, every rate carries the $s_{13}$-suppressed factor $|U_{e3}|^2\approx0.022$; the spectrum is minimally distorted. This produces the strongest NO--IO asymmetry among the three flavor-specific structures (see figure~\ref{fig:dsnb_std_gee}).
    \item \textit{Muon-specific coupling}~($g_{\mu\mu}$): The dependence on the mass ordering is inverted as compared to $e-$specific case. In NO, absorption is suppressed by $|U_{\mu 1}|^2\approx0.07$, resulting in a weaker depletion effect, whereas in IO, $|U_{\mu 3}|^2\approx0.56$ leads to substantial suppression in all flavors (see figure~\ref{fig:dsnb_std_gmm}).
    \item \textit{Tau-specific coupling}~($g_{\tau\tau}$): In NO, $|U_{\tau 1}|^2\approx0.25$ leads to a moderate effect, while in IO, $|U_{\tau 3}|^2\approx0.42$ gives comparatively stronger suppression. Since the NO and IO weights are roughly similar in magnitude, the ordering dependence is the weakest among the three flavor-specific scenarios (see figure~\ref{fig:dsnb_std_gtt}).
\end{enumerate}

\section{\label{events}Detection Prospects at Neutrino Observatories}
The detection of DSNB signal requires neutrino detectors with large fiducial volumes and strong background suppression in the relevant energy window. The next-generation experiments HK-Gd and DUNE, together with the recently commissioned JUNO, are expected to provide the required sensitivity. In this section, we compute the expected DSNB event rates at these detectors and quantify how the $\nu$SI-modified fluxes translate into observable spectral distortions above the relevant backgrounds.

\subsection{JUNO}\label{exp:juno}
The Jiangmen Underground Neutrino Observatory (JUNO) is a 20~kton liquid scintillator detector located in Guangdong province, China~\cite{JUNO:2015zny} with an overburden of 700~m of rock\footnote{See ref.~\cite{JUNO:2025gmd} for the initial results from JUNO using reactor neutrinos.}. The primary detection channel for the DSNB is inverse beta decay (IBD), $\bar{\nu}_e+p\rightarrow e^+ +n$. The main backgrounds affecting the DSNB energy window include reactor $\bar\nu_e$ (dominant below $\sim12$ MeV), atmospheric neutrino neutral current (NC) interactions on $^{12}$C, atmospheric charged current (CC) $\nu_e$ events (dominant above $\sim30$ MeV), fast neutrons, and cosmogenic $^9$Li/$^8$He. The application of pulse shape discrimination (PSD) techniques~\cite{Cheng:2023zds}, fiducial volume cuts, and muon vetoes can substantially suppress these backgrounds, defining an optimal signal window of 12--30 MeV in prompt energy. We adopt the background estimates and detection efficiencies from ref.~\cite{JUNO:2022lpc}\footnote{ We follow the updated analysis of ref.~\cite{JUNO:2022lpc}, which improves upon ref.~\cite{JUNO:2015zny} with more realistic backgrounds and optimized PSD efficiency.}.

\noindent The expected events per energy bin can be calculated as
\begin{equation}\label{eq:JUNO_evtrate}
    N_{i} = \xi_{\rm JUNO}\cdot N_{\rm JUNO}\cdot T_{\rm JUNO}\,\int_{E_{e^+}^{i,~min}}^{E_{e^+}^{i,~max}}dE_{e^+}\int  dE_{e^+}^{'}~\mathcal{R}(E_{e^+},E_{e^+}^{'})~\Phi_{\bar{\nu}_e }(E_\nu)~\sigma_{\rm IBD}(E_{\nu}),
\end{equation}

\begin{figure}[!t]
    \centering
    \includegraphics[width=0.495\linewidth]{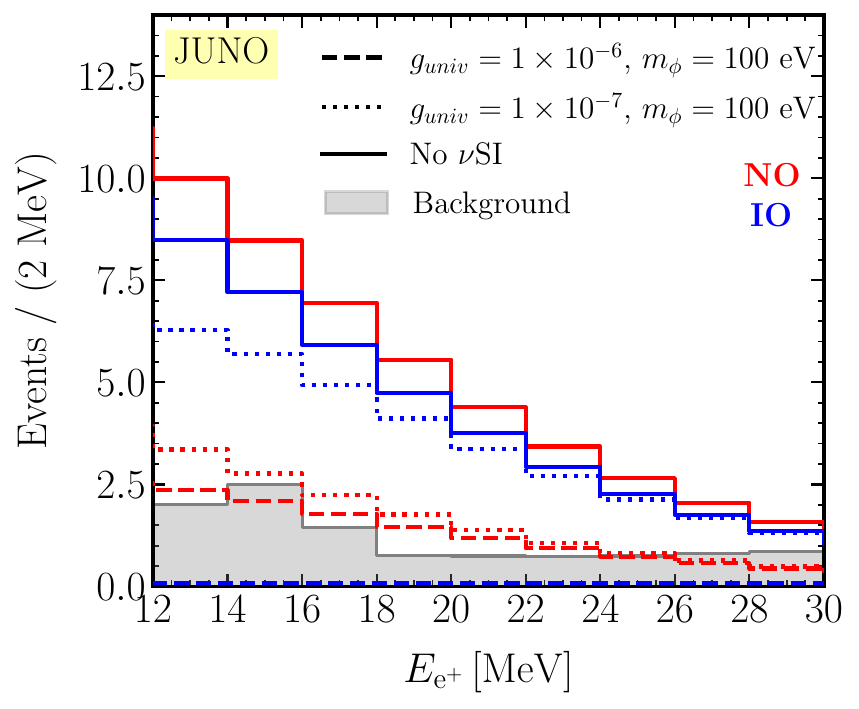}
    \includegraphics[width=0.495\linewidth]{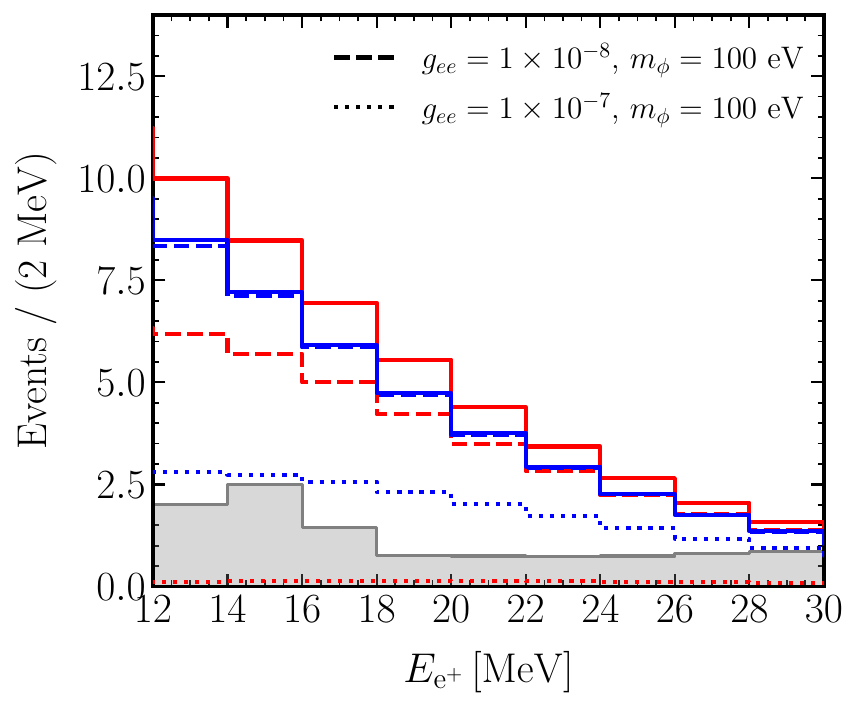}
    \includegraphics[width=0.495\linewidth]{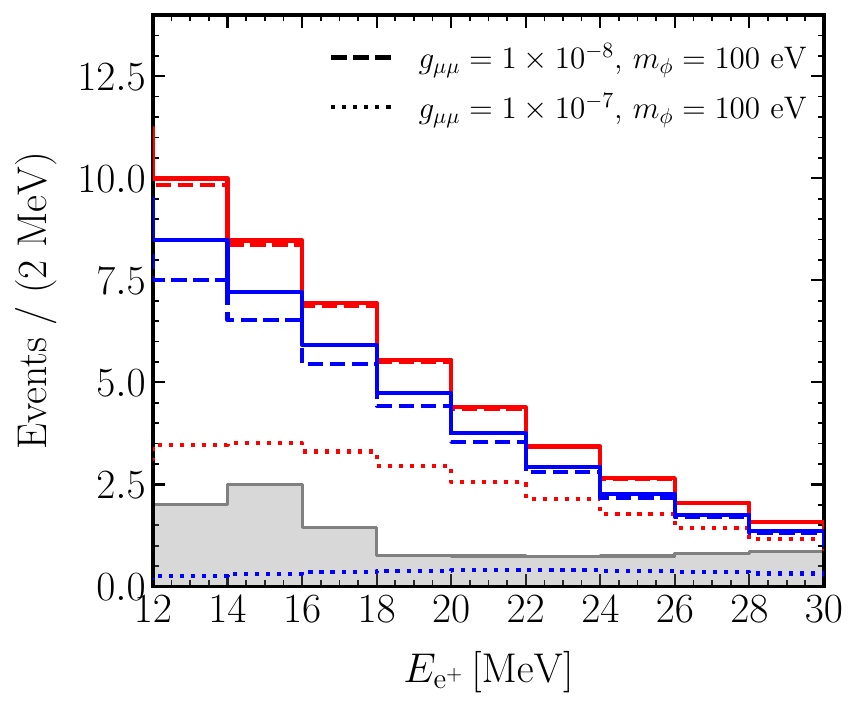}
    \includegraphics[width=0.495\linewidth]{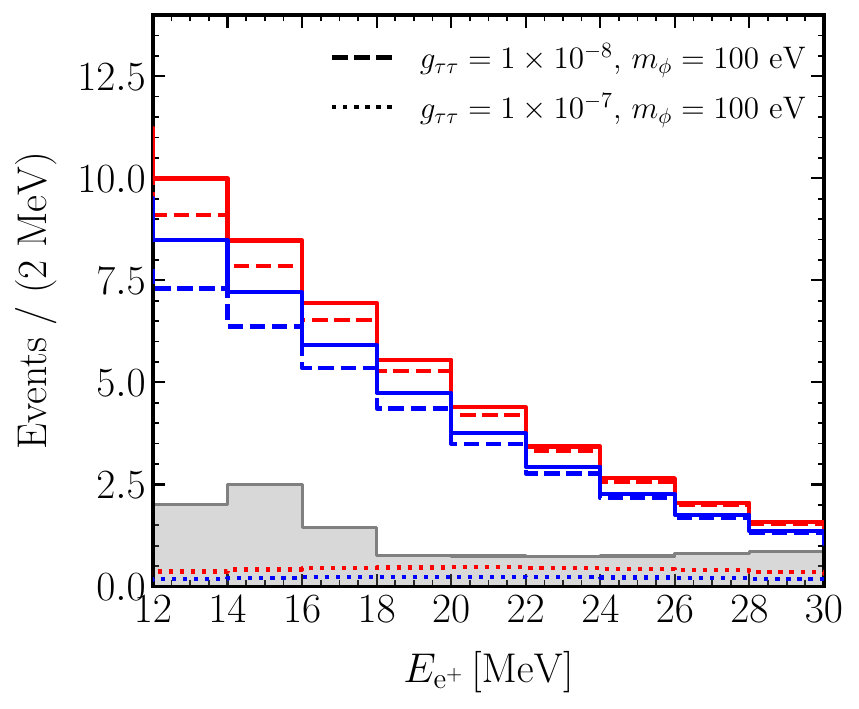}
    \caption{Projected event rates at JUNO for 20 years of exposure with an 18 kton fiducial volume. The solid lines show the free-streaming scenario, while the dashed and dotted lines show the $\nu$SI-modified event rates for two representative coupling values with $m_\phi= 100$~eV. The red and blue correspond to NO and IO, respectively. The gray shaded region denotes the total background. The four panels assume universal (top left), $e$-specific (top right), $\mu$-specific (bottom left), and $\tau$-specific (bottom right) coupling structures. The distinct suppression patterns across panels and orderings reflect the PMNS weighting of the mass-basis couplings $|g_{ik}|^2$ (see text for discussion).}
    \label{fig:JUNO_evtrate}
\end{figure}

where $N_{i}$ is the number of events in the $i-$th energy bin, $\Phi_{\bar{\nu}_e }(E_\nu)$ is the $\bar\nu_e$ flux at Earth, obtained from the oscillation-averaged projection of eq.~(\ref{eq:flavproj}), using either the free-streaming or the $\nu$SI-modified mass-basis fluxes from section~\ref{sec:nusi}. Here, $E_{\nu}$ is the true neutrino energy, while $E_{e^+}'$ and $E_{e^+}$ represent the true and reconstructed positron energies, respectively. The positron energy is related to the neutrino energy by $E_{e^+}'= E_\nu-\Delta$, with $\Delta = m_n - m_p \approx 1.293$, and $\sigma_{\rm IBD}(E_\nu)$ is the differential IBD  cross-section~\cite{Strumia:2003zx} which depends on the incident neutrino energy. The parameters $T_{\rm JUNO}$, $N_{\rm JUNO}$, and $\xi_{\rm JUNO}$ denote the exposure time, the number of free proton targets, and the detector efficiency, respectively. Following ref.~\cite{JUNO:2022lpc}, we assume a fiducial mass of approximately 18 kton---combining 14.7 kton inner (FV1) and 3.6 kton outer (FV2) detector volumes, a runtime of 20 years, and average detection efficiencies of 84\% and 77\% for the inner and outer volumes, respectively. The number of free proton targets is $7.15\times10^{31}$ kton$^{-1}$. The detector response function $\mathcal{R}(E_{e^+}, E_{e^+}')$ is modeled as a Gaussian profile,

\begin{equation}\label{eq:Gauss_resolution}
    \mathcal{R}(E_{e^+},E_{e^+}^{'})=\frac{1}{\delta_E\sqrt{2\pi}}\exp\Big[-\frac{(E_{e^+}-E_{e^+}^{'})^2}{2\delta_{E}^2}\Big]
\end{equation}
with $\delta_E/E_{e^+} =3\%/\sqrt{E_{e^+}/{\rm MeV}}$~\cite{JUNO:2015zny}. We adopt an energy bin width of 2~MeV to ensure sufficient statistics in each energy bin.

The projected JUNO event rates for the four coupling structures are shown in figure~\ref{fig:JUNO_evtrate} with $m_\phi=100$~eV. For universal couplings (top-left panel), benchmark values $g_{\rm univ}=10^{-6}$~(dashed) and $10^{-7}$~(dotted) are shown; for flavor-specific couplings, $g_{\alpha\alpha}=10^{-8}$~(dashed) and $10^{-7}$~(dotted) are used. Since JUNO detects $\bar\nu_e$ via IBD, the event spectrum directly reflects the $\nu$SI-modified $\Phi_{\bar\nu_e}$ discussed in section~\ref{sec:mod_spectra}, convolved with the IBD cross section and detector response. The ordering-dependent suppression patterns for each coupling structure follow the same logic summarized there---in particular, the coupling-dependent NO--IO crossover in the universal case is directly visible in the top-left panel.

\subsection{Hyper-Kamiokande with Gadolinium}\label{exp:hkgd}
Hyper-Kamiokande (HK) is a proposed next-generation water Cherenkov neutrino detector currently under construction in Japan, with a fiducial volume of 187~kton per tank~\cite{Hyper-Kamiokande:2018ofw}. Similar to JUNO, IBD serves as the primary detection channel for HK. We assume that HK is loaded with 0.1\% $\mathrm{GdCl}_3$ in water, which enables efficient neutron tagging~\cite{Nairat:2024upg} that suppresses the invisible muon background by approximately a factor of five and thereby improves DSNB sensitivity~\cite{Hyper-Kamiokande:2018ofw}. Beyond invisible muon backgrounds, the relevant backgrounds in the DSNB energy range include cosmogenic $^9$Li spallation products, atmospheric CC $\nu_e/\bar\nu_e$ events which dominate above $\sim$20~MeV, and atmospheric NC quasi-elastic interactions with nuclei that produce de-excitation $\gamma$-rays and neutrons that can mimic the IBD signature. The impact of these backgrounds can be substantially reduced through muon vetoes and efficient neutron tagging~\cite{Nairat:2024upg}. Following refs.~\cite{Akita:2022etk,Balantekin:2023jlg}, we neglect the NC background, assuming it can be efficiently suppressed using machine learning-based discrimination techniques~\cite{Maksimovic:2021dmz,Moller:2018kpn}. Further improvements in background rejection and neutron-tagging performance may additionally enhance the DSNB sensitivity at HK. We adopt the background estimates from the HK design report~\cite{Hyper-Kamiokande:2018ofw,Huang:2015hro}\footnote{The background components included in our analysis are consistent with those considered in refs.~\cite{Moller:2018kpn, Balantekin:2023jlg}} and assume an optimal signal window of 12--30~MeV in reconstructed positron energy. 

The expected event rate for HK-Gd is computed by following the eq.~\eqref{eq:JUNO_evtrate} but with HK-Gd detector configurations: two tanks with a total fiducial volume of 374~kton ($N_{\rm HK} = 2.5\times10^{34}$ free proton targets), a runtime of $T_{\rm HK}=$10~years (corresponding to an exposure of 3740~kton$\cdot$yr)\footnote{Equivalently, 20 years of single-tank (187~kton) operation.}, and an overall detection efficiency of $\xi_{\rm HK} \approx 67\%$ -- combining 90\% neutron capture efficiency with 74\% event selection efficiency. The detector response is modeled using a Gaussian profile as shown in eq.~\eqref{eq:Gauss_resolution}, with an energy resolution of
\begin{equation}
    \delta_E/E_{e^+} = \frac{0.3162}{\sqrt{E_{e^+}/\mathrm{MeV}}} - \frac{0.05525}{E_{e^+}/\mathrm{MeV}} + 0.04572\,,
\end{equation}
assuming performance comparable to Super-Kamiokande~\cite{Super-Kamiokande:2023jbt}, with only modest expected improvements~\cite{Super-Kamiokande:2025sxh}.
\begin{figure}[!h]
    \centering
    \includegraphics[width=0.495\linewidth]{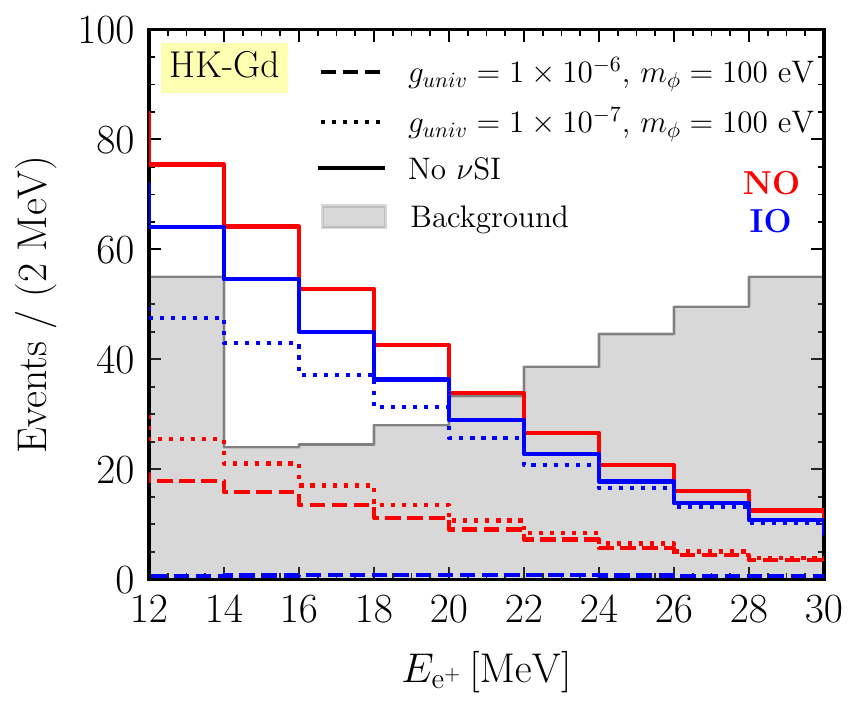}
    \includegraphics[width=0.495\linewidth]{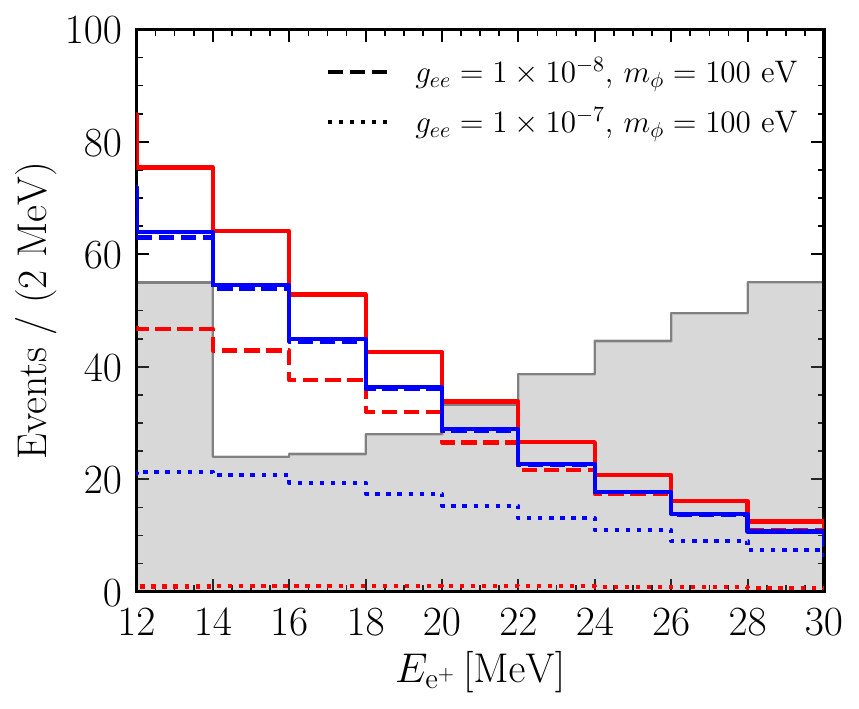}
    \includegraphics[width=0.495\linewidth]{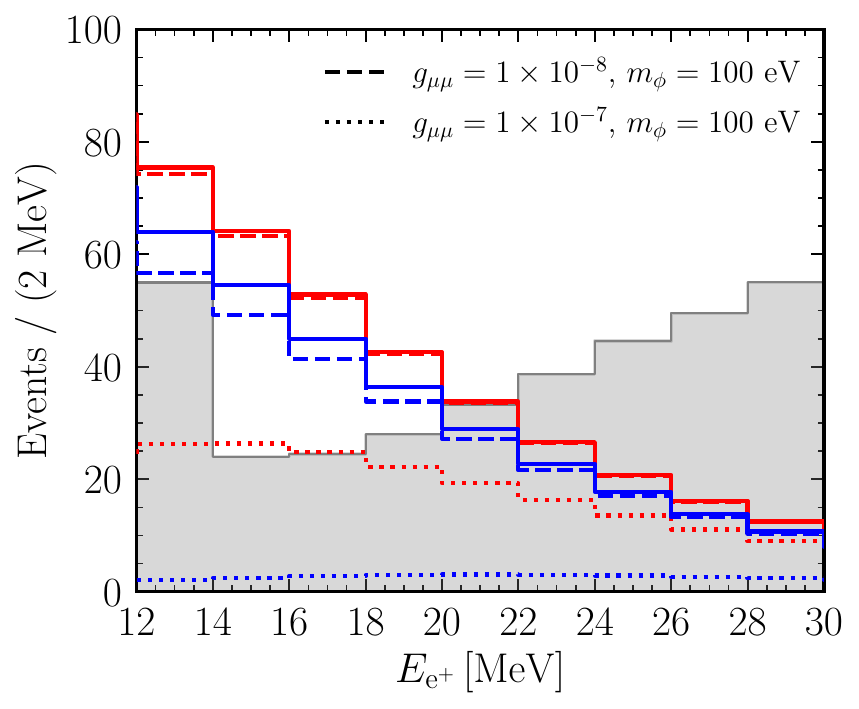}
    \includegraphics[width=0.495\linewidth]{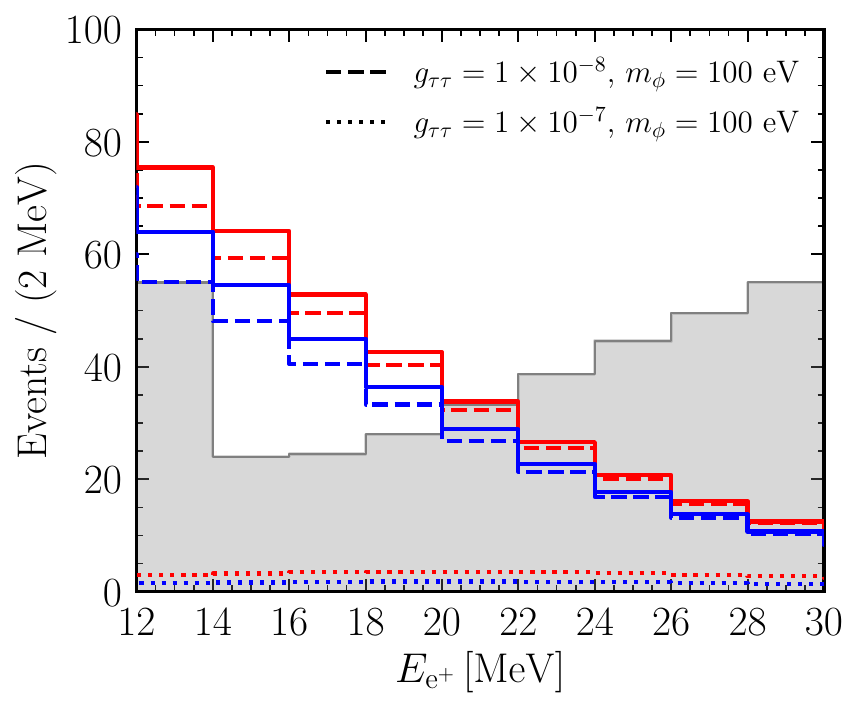}
    \caption{Same as figure~\ref{fig:JUNO_evtrate}, but for HK-Gd assuming 10 years of operation with a total fiducial volume of 374 kton. The larger target mass yields higher event counts, improving the statistical discrimination between the free-streaming and $\nu$SI-modified spectra across all four coupling structures.}
    \label{fig:HK_evtrate}
\end{figure}

In figure~\ref{fig:HK_evtrate}, we show the projected HK-Gd event rates for the same benchmark couplings as those used for JUNO, with $m_\phi=100$~eV. Since HK also detects $\bar\nu_e$ via IBD, the qualitative suppression patterns for each coupling structure are the same as at JUNO (see section~\ref{exp:juno}). The larger fiducial volume increases the signal statistics by nearly an order of magnitude, improving the ability to distinguish the free-streaming and $\nu$SI-modified spectra. However, backgrounds also scale with detector mass, leading to a less favorable signal-to-background ratio than at JUNO, especially above $\sim$20~MeV.

\subsection{DUNE}\label{exp:dune}
The Deep Underground Neutrino Experiment (DUNE) is an upcoming long-baseline neutrino experiment that will employ a 40~kton liquid argon time projection chamber at the Sanford Underground Research Facility in South Dakota, USA~\cite{DUNE:2020ypp}. Unlike JUNO and HK, which detect $\bar\nu_e$ via IBD, DUNE is sensitive to $\nu_e$ through the charged-current interaction
\begin{equation}\label{eq:DUNE_Ar}
    \nu_e + {}^{40}\mathrm{Ar} \rightarrow e^- + {}^{40}\mathrm{K}^*\,,
\end{equation}
\begin{figure}[!b]
    \centering
    \includegraphics[width=0.495\linewidth]{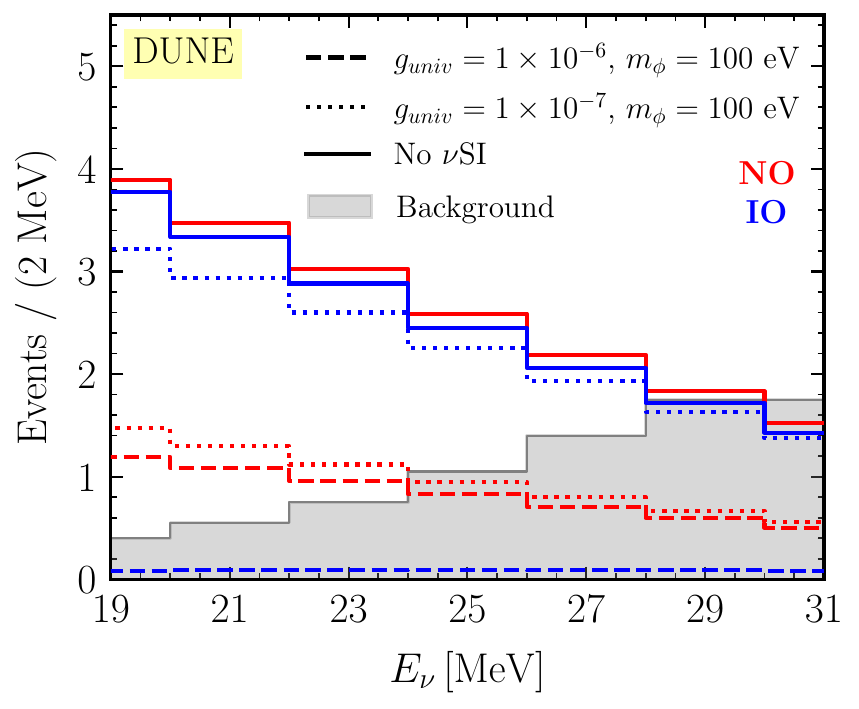}
    \includegraphics[width=0.495\linewidth]{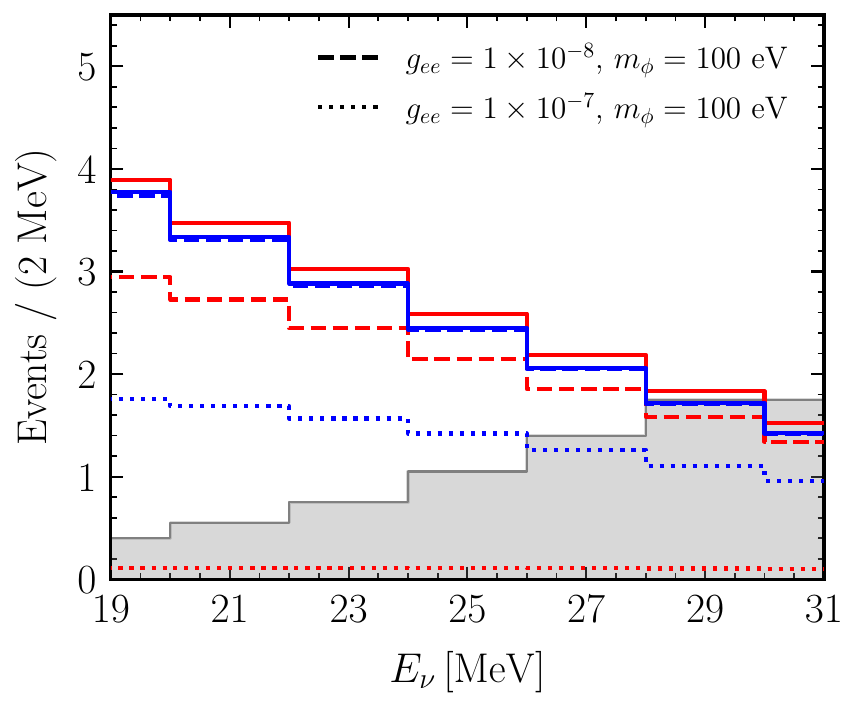}
    \includegraphics[width=0.495\linewidth]{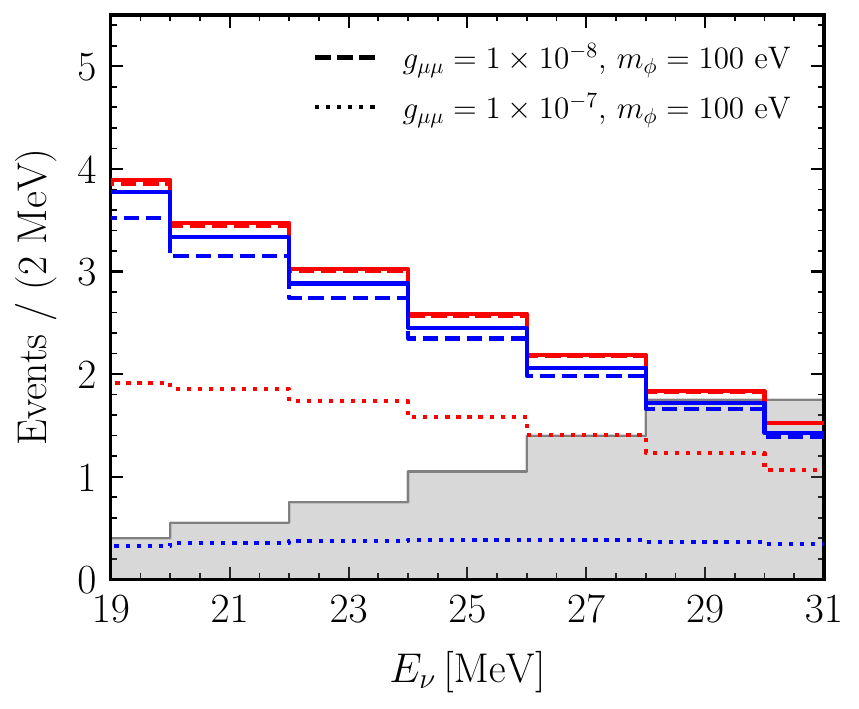}
    \includegraphics[width=0.495\linewidth]{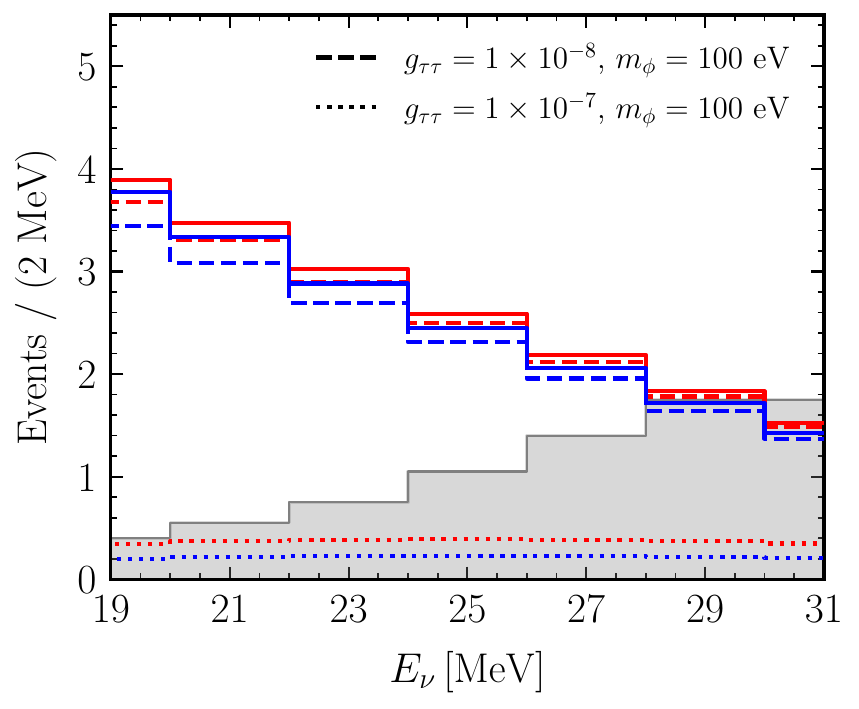}
    \caption{Same as figure~\ref{fig:JUNO_evtrate}, but for DUNE assuming 20~years of operation with a 40~kton fiducial volume. The optimal signal window spans 19--31~MeV. DUNE detects $\nu_e$ via CC interactions on $^{40}\mathrm{Ar}$, probing a different combination of mass eigenstates than the IBD channel at JUNO and HK, and hence exhibiting a distinct pattern of ordering-dependent suppression across coupling structures.}
    \label{fig:DUNE_evtrate}
\end{figure}
providing direct access to the $\nu_e$ component of the DSNB flux and thereby complementing the $\bar\nu_e$ measurements at JUNO and HK. Collectively, these experiments can provide sensitivity to the flavor-dependent effects of $\nu$SI on the DSNB. The dominant backgrounds in the low-energy regime are \emph{hep} and ${}^8$B solar neutrinos, which are significant below $\sim$19~MeV. At higher energies, irreducible atmospheric $\nu_e$ CC events become increasingly dominant. These considerations define an optimal signal window of 19--31~MeV. Although the background contributions and detector efficiency for low-energy DSNB searches remain under study, we follow the background treatment presented in ref.~\cite{Cocco:2004ac}.

The expected event rate follows the same calculation as eq.~\eqref{eq:JUNO_evtrate}, with $\Phi_{\bar\nu_e}$ replaced by the $\nu_e$ flux $\Phi_{\nu_e}(E_\nu)$ and the IBD cross section replaced by the $\nu_e$--${}^{40}\mathrm{Ar}$ CC cross-section $\sigma_{\nu_e\text{-Ar}}(E_\nu)$~\cite{GilBotella:2003sz}. We assume a fiducial mass of 40 kton ($N_{\rm DUNE} = 6.02\times10^{32}$ target atoms), a runtime of $T_{\rm DUNE}=$20~years, and an overall detection efficiency of $\xi_{\rm DUNE} \approx 86\%$ (combining 90\% trigger efficiency with 96\% reconstruction efficiency~\cite{Ankowski:2016lab}). Using the detector response described in eq.~\eqref{eq:Gauss_resolution}, we assume an energy resolution of $\delta_E/E_\nu = 20\%$~\cite{DUNE:2020ypp}. While this resolution is substantially poorer than JUNO’s and consequently washes out sharp spectral features, the broad absorption signature associated with the relativistic-C$\nu$B regime remains visible.

The projected DUNE event rates are shown in figure~\ref{fig:DUNE_evtrate} for the same benchmark couplings and $m_\phi=100$~eV. Because DUNE detects $\nu_e$ via CC absorption on argon rather than $\bar\nu_e$ via IBD, it probes a different MSW branch: the $\nu_e$ emission maps onto $\nu_3$ in NO and $\nu_2$ in IO, as opposed to the $\bar\nu_e\to\bar\nu_1$~(NO) and $\bar\nu_e\to\bar\nu_3$~(IO) mapping relevant for JUNO and HK. The direction of the ordering-dependent suppression for the flavor-specific structures follows the same logic discussed in section~\ref{sec:mod_spectra}, since the PMNS weights $|U_{e i}|^2$ entering $\Phi_{\nu_e}$ are the same as those entering $\Phi_{\bar\nu_e}$. For universal couplings, the coupling-dependent NO--IO crossover discussed there applies equally to the DUNE event rates.

\section{\label{res}Projected Sensitivity to Secret Neutrino Interactions}
We now derive the projected sensitivity of JUNO, HK-Gd, and DUNE to $\nu$SI in the $(m_\phi,\,g)$ parameter plane for each of the four coupling structures. To quantify the statistical significance, we adopt the Asimov approach and construct the statistical $\chi^2$ as
\begin{equation}\label{eq:chi2_fn}
\chi^{2} = \min_{y}\left\{ 2\sum_{i}\left[ \mu_{i}(y) - n_{i} + n_{i}\ln\frac{n_{i}}{\mu_{i}(y)} \right] + \left(\frac{y}{\sigma_{y}}\right)^{2} \right\},
\end{equation}
\noindent where $\mu_i(y) = (1+y)N_i^{\nu\mathrm{SI}} + N_i^{\mathrm{bkg}}$ and $n_i = N_i^{\mathrm{FS}} + N_i^{\mathrm{bkg}}$ denote the event counts in the $i$-th energy bin. The quantities $N_i^{\nu\mathrm{SI}}$ and $N_i^{\rm FS}$ represent the number of events in the $i$-th energy bin under the $\nu$SI and free-streaming scenario, respectively, and $N_i^{\rm bkg}$ represents the contribution from background events. The nuisance parameter $y$ is a normalization factor with an associated uncertainty of $\sigma_{y}=0.5$, introduced to account for theoretical uncertainties in the DSNB flux arising from the core-collapse rate, the progenitor spectrum, and the black-hole formation fraction~\cite{Wang:2025qap}. We incorporate the nuisance parameter following the pull method and minimize the $\chi^2$ over $y$.

We derive the $3\sigma$ exclusion contours from a test statistic defined as $\Delta\chi^2\equiv\chi^2-\chi^2_{\min}$\footnote{In the Asimov treatment, no statistical fluctuations are considered such that $\chi^2_{\min} = 0$.}. For two degrees of freedom, the $3\sigma$ contours correspond to $\Delta \chi^2 = 11.83$. The experimental specifications adopted for each detector are summarized in table~\ref {tab:DSNB_experiments}.
\begin{table}[!h]
\centering
\caption{Detector configurations adopted in the sensitivity analysis: fiducial volume, runtime, detection efficiency ($\xi$), and optimal energy window.}
\begin{tabular}{lcccc}
\toprule
Experiment & Fiducial volume & Runtime & Detection efficiency & Energy window \\
& {[kton]} & {[years]} & {$\xi~(\%)$} & {[MeV]} \\
\midrule
JUNO ~($\bar{\nu}_e$) & FV1(14.7)~$\oplus$~FV2(3.6) &  20 & 84~$\oplus$~77 & 12--30 \\
HK-Gd ($\bar{\nu}_e$) & 374 & 10 &  67 & 12--30 \\
DUNE ~(${\nu}_e$) & 40 & 20 &  86 & 19--31 \\
\bottomrule
\end{tabular}
\label{tab:DSNB_experiments}
\end{table}

\begin{figure}[!h]
    \centering
    \includegraphics[width=0.495\linewidth]{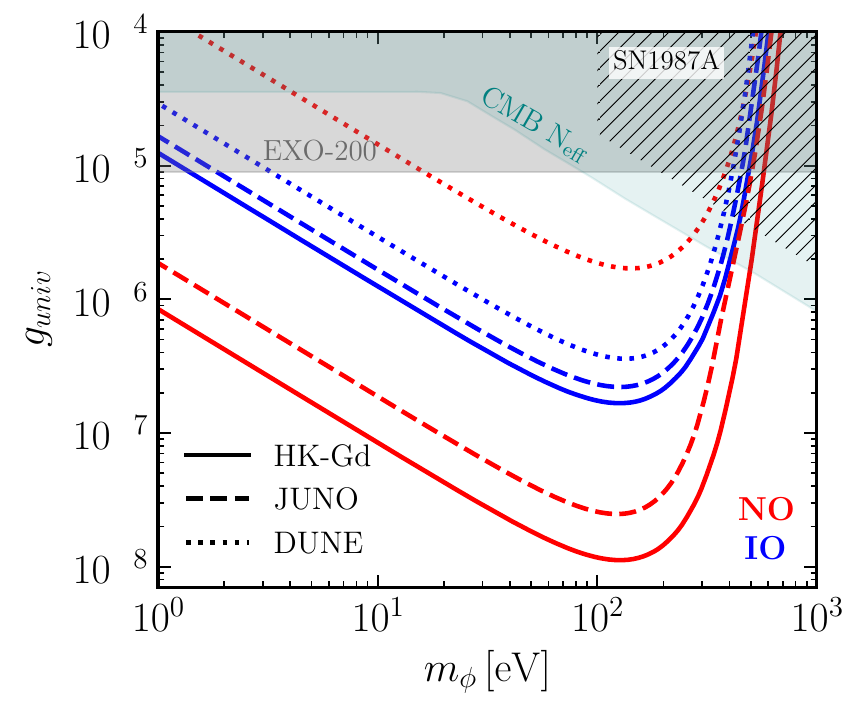}
    \includegraphics[width=0.495\linewidth]{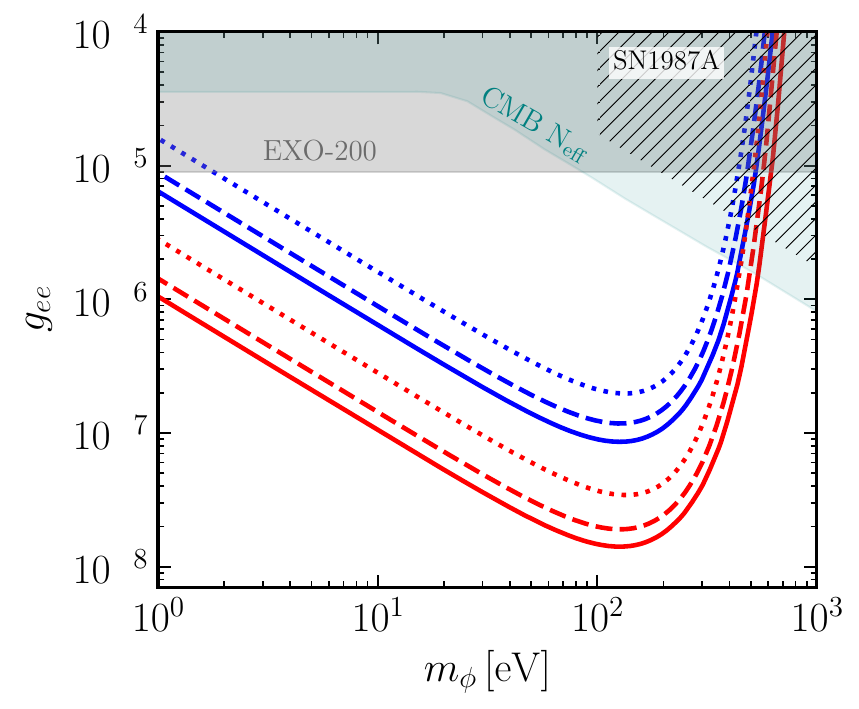}
    \includegraphics[width=0.495\linewidth]{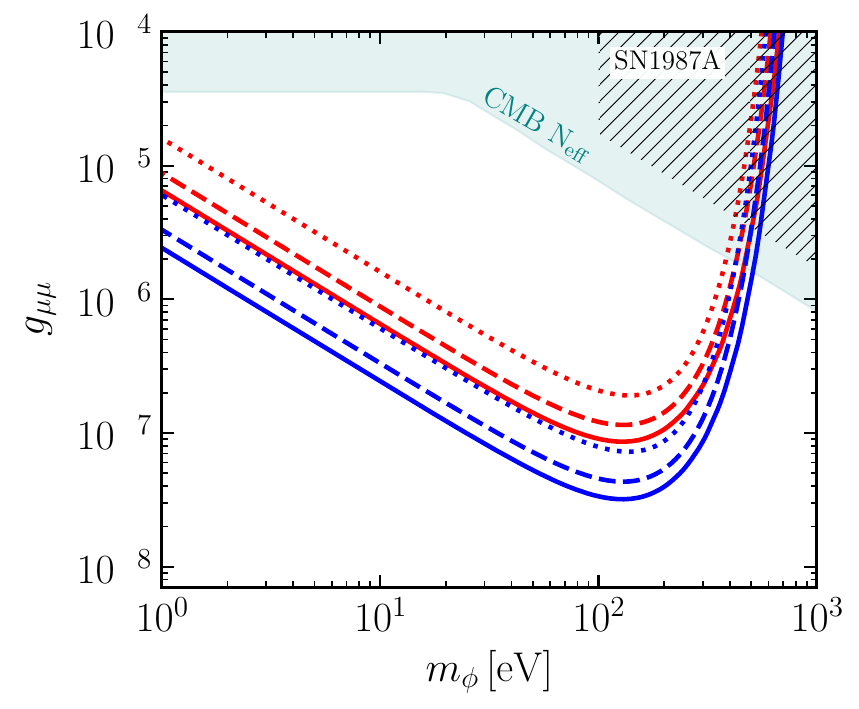}
    \includegraphics[width=0.495\linewidth]{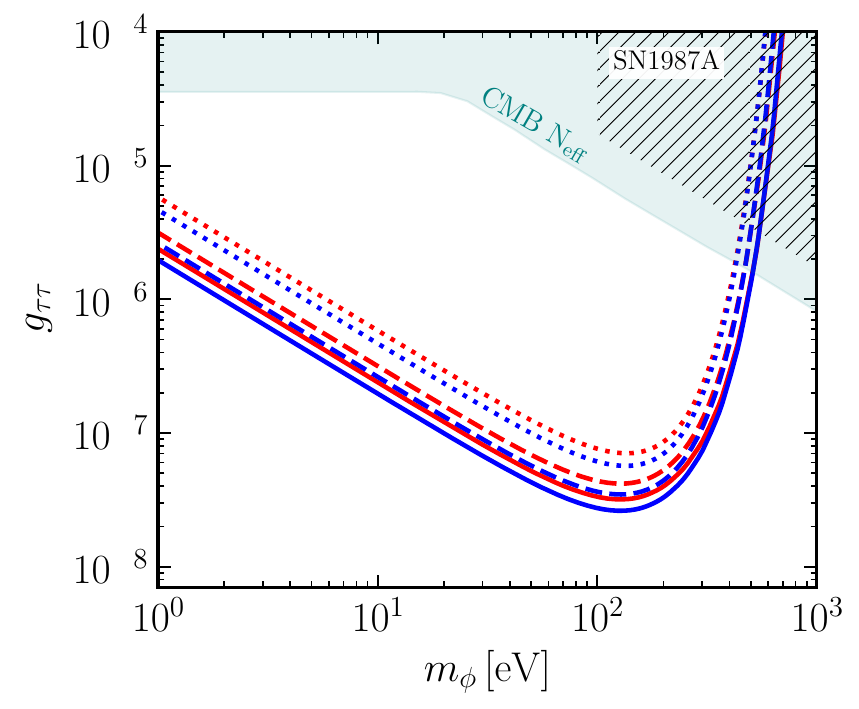}
    
    \caption{Projected $3\sigma$ exclusion contours in the $(m_\phi,\,g)$ plane for NO (red) and IO (blue). The solid, dashed, and dotted lines show the sensitivities for HK-Gd, JUNO, and DUNE, respectively. The four panels correspond to universal (top left), $e$-specific (top right), $\mu$-specific (bottom left), and $\tau$-specific (bottom right) coupling structures. Shaded regions denote existing constraints: $0\nu\beta\beta\phi$ from EXO-200~\cite{Kharusi:2021jez} (applicable to $g_{\rm univ}$ and $g_{ee}$ only), CMB $N_{\rm eff}$~\cite{Li:2023puz}, and SN~1987A~\cite{Fiorillo:2022cdq}. The latter two were obtained for flavor-universal couplings and are shown unchanged in the flavor-specific panels, where they should be read as indicative (see text).
    }
    \label{fig:sensitivity}
\end{figure}

In figure~\ref{fig:sensitivity}, we show the projected $3\sigma$ sensitivity contours in the $(m_\phi,\,g)$ parameter space for all four coupling structures, both mass orderings, and each of the three experiments. All contours exhibit a characteristic U-shape with optimal sensitivity at intermediate mediator masses ($m_\phi\sim100$--$300$~eV) and diminished reach at both ends. This reflects the overlap between the resonance energy and the detector signal window: the C$\nu$B-averaged resonance energy $E_\nu^{\rm res}\sim m_\phi^2/(4T_{\rm C\nu B})$ falls below the lower edge of the signal window for $m_\phi\lesssim$ few~eV, while for $m_\phi\gtrsim500$~eV it shifts into the steeply falling tail of the DSNB spectrum. The strongest reach is achieved when the absorption feature overlaps maximally with the region of the highest signal-to-background ratio.

For universal couplings (top-left panel), the ordering dependence reflects the coupling-dependent crossover discussed in section~\ref{sec:mod_spectra}. At moderate couplings the depletion of $\Phi_1$ in NO dominates the $\bar\nu_e$ ($\nu_e$) suppression via $|U_{e1}|^2\approx0.68$, while at larger couplings $\Phi_2$ remains protected in NO by its $s_{13}^2$-suppressed coupling to $\nu_1$, whereas in IO it is depleted at the $s_{13}$ level, causing IO to become more suppressed overall. Since HK-Gd and JUNO probe the smallest couplings, their contours lie in the regime where NO results in stronger sensitivity. The DUNE contours, which probe comparatively larger couplings, enter the regime where IO provides stronger sensitivity.

For the flavor-specific scenarios, the ordering dependence follows the patterns discussed in section~\ref{sec:mod_spectra}. The $e$-specific couplings result in the strongest NO--IO asymmetry, with IO sensitivity suppressed by $|U_{e3}|^2\approx0.022$ and pushed close to existing bounds. The $\mu$- and $\tau$-specific couplings invert the hierarchy, with IO providing stronger sensitivity. The $\tau$-specific case shows the least NO--IO separation due to the comparable magnitudes of $|U_{\tau 1}|^2\approx0.25$ and $|U_{\tau 3}|^2\approx0.42$. The strongest projected reach across all panels is achieved by HK-Gd in NO for universal couplings, where the contour extends close to $g_{\rm univ}\sim10^{-8}$ for $m_\phi\sim100-200$ eV. This is comparable to the reach quoted in ref.~\cite{Wang:2025qap}, although only for universal couplings and only in NO, since that work treats the DSNB as a single effective species and therefore represents an optimistic limiting case.\footnote{There the resonance depletes the entire flux, whereas here the depletion acts on individual mass eigenstates and reaches an IBD detector only through the $|U_{ei}|^2$ projection.}
Among the three experiments, HK-Gd (solid) provides the strongest sensitivity across all coupling structures, owing to its large fiducial volume. JUNO (dashed), despite a smaller target mass, benefits from a superior signal-to-background ratio and reaches sensitivities within a factor of $\sim$2--3 of HK across most of the parameter space. DUNE (dotted) provides independent and complementary sensitivity through the $\nu_e$ channel.

The shaded regions in figure~\ref{fig:sensitivity} correspond to the existing constraints. The $0\nu\beta\beta\phi$ bound from EXO-200~\cite{Kharusi:2021jez}, applicable only to the $g_{\rm univ}$ and $g_{ee}$ panels, excludes $g\gtrsim10^{-5}$ across the full mass range but lies well above the projected DSNB reach. The CMB $N_{\rm eff}$ bound~\cite{Li:2023puz}, derived from the dilution-resistant contribution of $\phi$ to the relativistic energy density, and the SN~1987A bound~\cite{Fiorillo:2022cdq} are more stringent for $m_\phi\gtrsim100$~eV. Below this scale, the SN~1987A analysis is not applicable and the $N_{\rm eff}$ bound weakens significantly. The DSNB therefore provides the leading projected sensitivity in this mass range, reaching couplings as low as $g\sim10^{-8}$. Moreover, the DSNB probes qualitatively distinct physics. It is flavor discriminating, in the sense that different coupling structures are weighted by distinct PMNS elements $|U_{\alpha k}|^2$, whereas the existing bounds have been derived for a flavor-universal coupling and are not straightforwardly recast for a specific flavor structure. A DSNB measurement would therefore allow identification of the underlying flavor structure of $\nu$SI if a signal is observed.

We caution that the CMB $N_{\rm eff}$ and SN~1987A curves in figure~\ref{fig:sensitivity} are plotted as given in refs.~\cite{Fiorillo:2022cdq,Li:2023puz}, so in the flavor-specific panels they should be read as indicative only. Neither maps onto a flavor-specific coupling by a simple rescaling. The SN~1987A limit depends on the coupled flavor both through production in the core and through the flavor composition observed at the detector, while the $N_{\rm eff}$ limit depends on the number of coupled species in a way that is not a simple overall factor, since the relevant production channels scale differently with it. On general grounds we expect the SN~1987A limit to be weaker for $\mu$- and $\tau$-specific couplings and to change little for $e$-specific ones, and the $N_{\rm eff}$ limit to be weaker than the curve shown in all three flavor-specific panels. A quantitative treatment is beyond the scope of this work.

\section{\label{conc} Conclusions}
With the advent of next-generation neutrino detectors, the observation of the DSNB is now within reach. JUNO has recently commenced data taking, while Hyper-Kamiokande and DUNE are expected to start data collection by the end of this decade. Each experiment offers distinct advantages, such as JUNO benefits from its exceptional energy resolution, Hyper-Kamiokande from its large event statistics, and DUNE from its sensitivity to $\nu_e$ flavor, providing complementary information to the global DSNB search. The observed DSNB spectra will carry signatures of both the core-collapse population integrated over cosmic history and any non-standard neutrino interactions encountered during the propagation of neutrinos over cosmological distances through the C$\nu$B. In this work, we explore how $\nu$SI mediated by a light scalar $\phi$ can modify the DSNB flux and the observable event rates at next-generation neutrino detectors, providing sensitivity projections in the $(m_\phi ,\ g)$ parameter space for different coupling scenarios.

This work extends the analysis of ref.~\cite{Wang:2025qap} in four essential aspects. Firstly, we employ a full three-flavor framework using the complete PMNS mixing matrix, including non-zero $\theta_{13}$, non-maximal $\theta_{23}$, and a non-zero but fixed Dirac CP phase $\delta_{CP}$. Such a treatment captures the mass ordering and flavor-dependent phenomenology. It also reveals features such as the $\mu$--$\tau$ flux splitting, which is absent under the $\theta_{13}=0$ or maximal-$\theta_{23}$ approximations. Secondly, we present results for both normal and inverted mass orderings, identifying distinctive absorption signatures that could in principle discriminate between them. Third, we explore four representative flavor-diagonal coupling structures---universal, $e$-, $\mu$-, and $\tau$-specific that span the qualitatively distinct phenomenology accessible to scalar-mediated $\nu$SI. Fourth, we solve the coupled seven-component Boltzmann system that separately tracks the three neutrino and three antineutrino mass eigenstates, capturing the initial particle--antiparticle asymmetry of the DSNB source spectra together with the mediator dynamics.

The resonant scattering $\nu_i\nu_k\to\phi$ off the lightest relativistic C$\nu$B state produces a broad spectral depletion in the DSNB flux. The depletion pattern across mass eigenstates is dictated by the mass-basis coupling $g_{ik}$, while the projection onto observed flavors is governed by the weights $|U_{\alpha i}|^2$. This interplay leads to distinct, mass-ordering-dependent signatures for each coupling structure, which are directly seen in the event spectra of all the experiments considered in this work (figures~\ref{fig:JUNO_evtrate}--\ref{fig:DUNE_evtrate}).

For mediator masses $m_\phi\sim100$--$300$~eV, the resonance energy coincides with the detector signal window, which enables a projected $3\sigma$ sensitivity to couplings as small as $g\sim10^{-8}$, as illustrated in figure~\ref{fig:sensitivity}. Among the three experiments, HK-Gd provides the strongest reach owing to its high statistics, with JUNO within a factor of $\sim$2--3 across most of the parameter space due to its better signal-to-background ratio, and DUNE providing independent and complementary sensitivity through the $\nu_e$ channel. Crucially, in the sub-100~eV mass range, the DSNB surpasses all existing bounds, including CMB $N_{\rm eff}$, SN~1987A, and $0\nu\beta\beta\phi$ by up to a few orders of magnitude. Even in the parameter space where the CMB $N_{\rm eff}$ bound is formally competitive, the DSNB offers \emph{flavor-discriminating} information. A synergistic approach that exploits the complementary capabilities of different experiments may further strengthen the sensitivity to such new physics effects beyond the reach of individual experiments.

\acknowledgments
We are grateful to Xun-Jie Xu for helpful discussions. This work is financially supported by the Indian Association for the Cultivation of Science (IACS), Kolkata.

\bibliographystyle{jhep}
\bibliography{bib_nSI}

\end{document}